\documentclass[reqno]{amsart}

\usepackage{amsmath}
\usepackage{amssymb}
\usepackage{amsthm}
\usepackage{bm}
\usepackage{color}
\usepackage{eucal}
\usepackage{fullpage}
\usepackage{graphicx}
\usepackage{hhline}
\usepackage{lineno}
\usepackage[sc]{mathpazo}
\usepackage{mathrsfs}
\usepackage{mathtools}
\usepackage[numbers,sort&compress,square]{natbib}
\usepackage{setspace}
\usepackage{subfig}
\usepackage{xr}
\usepackage[all,cmtip]{xy}

\newcommand*\patchAmsMathEnvironmentForLineno[1]{%
	\expandafter\let\csname old#1\expandafter\endcsname\csname #1\endcsname
	\expandafter\let\csname oldend#1\expandafter\endcsname\csname end#1\endcsname
	\renewenvironment{#1}%
	{\linenomath\csname old#1\endcsname}%
	{\csname oldend#1\endcsname\endlinenomath}}%
\newcommand*\patchBothAmsMathEnvironmentsForLineno[1]{%
	\patchAmsMathEnvironmentForLineno{#1}%
	\patchAmsMathEnvironmentForLineno{#1*}}%
\AtBeginDocument{%
	\patchBothAmsMathEnvironmentsForLineno{equation}%
	\patchBothAmsMathEnvironmentsForLineno{align}%
	\patchBothAmsMathEnvironmentsForLineno{flalign}%
	\patchBothAmsMathEnvironmentsForLineno{alignat}%
	\patchBothAmsMathEnvironmentsForLineno{gather}%
	\patchBothAmsMathEnvironmentsForLineno{multline}%
}

\newcommand{\eq}[1]{Eq.~(\ref{eq:#1})}

\newcommand{\eqs}[2]{Eqs.~(\ref{eq:#1})-(\ref{eq:#2})}
\newcommand{\fig}[1]{Fig.~\ref{fig:#1}}
\newcommand{\figs}[2]{Figs.~\ref{fig:#1},~\ref{fig:#2}}

\newcommand{\app}[1]{\ref{app:#1}}

\begin{document}

\title{Asymmetric Evolutionary Games with Environmental Feedback}

\author{Christoph Hauert}
\author{Camille Saade}
\author{Alex McAvoy}

\begin{abstract}
Models in evolutionary game theory traditionally assume symmetric interactions in homogeneous environments. Here, we consider populations evolving in a heterogeneous environment, which consists of patches of different qualities that are occupied by one individual each. The fitness of individuals is not only determined by interactions with others but also by environmental quality. This heterogeneity results in asymmetric interactions where the characteristics of the interaction may depend on an individual's location. Interestingly, in non-varying heterogeneous environments, the long-term dynamics are the same as for symmetric interactions in an average, homogeneous environment. However, introducing environmental feedback between an individual's strategy and the quality of its patch results in rich eco-evolutionary dynamics. Thus, individuals act as ecosystem engineers. The nature of the feedback and the rate of ecological changes can relax or aggravate social dilemmas and promote persistent periodic oscillations of strategy abundance and environmental quality.
\end{abstract}

\maketitle

\section{Introduction}
Most models in evolutionary game theory tacitly assume that games are symmetric, which means that payoffs for individuals with distinct traits depend exclusively on their respective strategies and not on other features such as physical properties including size, strength, speed, and access to resources. Including any one of those aspects likely introduces at least small payoff differences, resulting in an asymmetric game \citep{maynard-smith:book:1982}. Since no two organisms are exactly alike in nature, such asymmetric interactions are actually the norm and symmetric games represent a (useful) idealization. Asymmetries can arise from genetic differences between individuals, from heterogeneity in environmental conditions, or from a combination of the two.

Actually, asymmetric interactions have long been recognized as important in evolutionary game theory, but they are usually addressed by considering interactions between two distinct populations \citep{taylor:JAP:1979,schuster:BiolCyber:1981}, or by assigning different roles to individuals, for example occupant and contender in territorial conflicts \citep{maynard-smith:AB:1976,hammerstein:AB:1981,gaunersdorfer:TPB:1991}. In the first case, the source of asymmetries is based on differences between species and not attributed to individual variation. Conversely, in the latter case, roles are typically assigned probabilistically, with interactions restricted to individuals in different roles. This precludes correlations between strategies and roles that would follow naturally from differences in individual features. Formally, this setup can be captured by symmetric interactions among individuals with a larger set of strategies that is augmented by the different roles \citep{sigmund:PNAS:2001}.

Of course, even asymmetric games are often oversimplifications of reality. It is also notoriously difficult to prove that reproductive fitness in nature can be explained in simple, game-theoretic terms \citep{turner:Nature:1999,nowak:Nature:1999,turner:AN:2003}. But these factors do not necessarily undermine the use of game theory in models. Indeed, in place of directly modeling the intricacies of real populations, evolutionary games provide a synthetic environment in which one can study the qualitative effects of \textit{some} features arising in natural populations, such as conflicts of interest and spatial heterogeneity. Given the traditional focus on symmetric games in these synthetic models, it is especially relevant to understand how sensitive the dynamics are to heterogeneity in payoff (fitness) accounting.

In the present work, we consider a natural source of asymmetry arising from heterogeneity in individual environments. In particular, we study evolutionary games where each player inhabits a ``patch,'' which can differ in quality and hence may account for location-dependent variation in reproductive fitness even without explicitly considering spatial dimensions. Over time, the quality of these patches can change through degradation or restoration, which introduces a temporal component to this spatial variation. As a result, the model naturally constitutes an evolutionary game with environmental feedback.

Including ecological components in evolutionary games has attracted increasing attention in recent years \citep{hauert:PRSB:2006,huang:PNAS:2015,mcnamara:JRSI:2013,gokhale:TPB:2016,weitz:PNAS:2016}. Spatial and temporal heterogeneity has also been treated extensively in population genetics, for example, when different alleles or traits are favored in different ecological niches \citep{levene:AN:1953,arnold:AN:1983} or if reproductive fitness depends on time \citep{haldane:JG:1963,ewing:AN:1979}. However, the population genetics literature has largely focused on the frequency-independent case and hence excludes evolutionary games. Heterogeneity in its different forms has long been recognized as a crucial part of evolution and frequency dependence represents a potent promoter.

In studies that do consider asymmetric, frequency-dependent interactions, the environment is typically considered to be fixed \citep{maynardsmith:AB:1976,selten:JTB:1980}. A standard example is the asymmetric replicator equation, which has been modified from its original form \citep{taylor:MB:1978} to include asymmetric payoffs and multiple populations \citep{hofbauer:JMB:1996,hofbauer:CUP:1998,hofbauer:BAMS:2003,accinelli:DGS:2011}. The replicator equation has also been extended to structured populations for both symmetric \citep{ohtsuki:JTB:2006} and asymmetric \citep{mcavoy:PLoSCB:2015} matrix games.

In finite populations, asymmetric games are usually presented in the form of bi-matrix games, where the fixation probability of a rare strategic type provides a measure of its evolutionary success \citep{ohtsuki:JTB:2010,sekiguchi:DGA:2015}. Bi-matrix games capture what some authors refer to as ``truly asymmetric" games \citep{cressman:PNAS:2014}, which, for finitely many individuals, could describe interactions between separate populations \citep{veller:JET:2016} or interactions along an edge on an evolutionary graph.

We show here that, in addition to affecting the evolutionary dynamics, environmental feedback can also introduce interesting ambiguities about the underlying nature of the strategic traits. For example, when cooperators benefit both co-players and environment, this trait can be considered ``cooperation" in an absolute sense. However, when cooperators must exploit their environment in order to provide benefits to co-players (such as through the extraction of valuable resources), this trait is only ``cooperation" relative to the co-player; to the environment, it better resembles defection. This ambiguity could be considered a behavioural equivalent to pleiotropy in genetics -- one behavioral strategy may simultaneously (and differently) impact the co-players' performance and the actor's environment. 

\section{Environmentally-induced asymmetric games}
In evolutionary models based on the replicator dynamics \citep{hofbauer:CUP:1998}, all that matters is the relative fitness of each strategic type compared to the others. In particular, whether fitness (or payoffs) are positive or negative has no meaning and no consequences on the dynamics. This property allows for some ambiguity in the interpretation of payoffs and (relative) fitness, and frequently the two terms are used interchangeably. On a more formal level, this ambiguity manifests itself in a convenient feature of the replicator equation, namely that the dynamics remain unchanged when adding a constant to any column of the payoff matrix \citep{hofbauer:CUP:1998}.

In the frequency-dependent Moran process \citep{nowak:Nature:2004}, conditions are more restrictive because fitness needs to be translated into transition probabilities and hence cannot be negative. Negative payoffs are usually resolved by mapping payoffs, $\pi$, to fitness, $f$, with common mappings of the form $f=\exp\left(w\pi\right)$ or $f=1+w\pi$, where $w$ indicates the selection strength, i.e. the contribution of the game payoffs to the overall fitness of individuals. If payoffs can be negative, the second form requires an upper limit on $w$ to ensure $f>0$. The important limit of weak selection, $w\ll 1$, ensures $f>0$, and actually renders the two payoff-to-fitness maps identical \citep{maciejewski:PLOSCB:2014}.

For mathematical convenience and tractability, most models in evolutionary game theory tacitly assume that the population is either infinite or is of finite but constant size. In contrast, ecological models emphasize the dynamics of a species' population size in a given environment or when competing with other species. In this case, fitness can be interpreted as the difference between the rates (or probabilities) of reproduction and death of individuals \citep{doebeli:eLife:2017}, which provides a viable interpretation for negative fitness; on the other hand, both the reproduction rate and the death rate need to be non-negative in order to remain biologically meaningful.

The purpose of this study is to incorporate ecological changes into evolutionary game theory by introducing feedback between an individual's strategic type and its environment. Interestingly, even though our model is based on the replicator dynamics, the interpretation of fitness requires more careful attention. Moreover, variable patch qualities introduce opportunities for ecosystem engineering \citep{jones:Oikos:1994,jones:Ecology:1997}. The lifespan of environmental changes can be ephemeral, like nests of passerine birds, or enduring, like beaver dams. When compared to the generation time of the species, the environmental changes determine the rate of eco-evolutionary feedback.

\subsection{Symmetric evolutionary games}
In the simplest formulation of an evolutionary game, individuals interact in pairs and can adopt, through imitation or inheritance, one of two strategic types. We refer to the two types as ``cooperate'' ($C$) and ``defect'' ($D$) because many interesting and relevant scenarios involve the problem of cooperation in social dilemmas \citep{dawes:ARP:1980}, i.e. in interactions that involve a conflict of interest between individual and joint performances. The most stringent and best-studied social dilemma is the prisoner's dilemma \citep{axelrod:book:1984} and its popular variant, the donation game \citep{sigmund:book:2010}.

In the donation game, a cooperator provides a benefit, $b$, to its partner at a cost to itself, $c$, where $0<c<b$. Defectors neither produce benefits nor incur costs. If both players cooperate, then both profit and obtain $b-c>0$; but both players are tempted to defect and avoid the costs. A defector meeting a cooperator gets $b$, whereas the cooperator is left with the costs, $-c$. Thus, a player is better off defecting regardless of what the opponent does; defection is the ``dominant'' strategy. However, if both players yield to temptation, they each end up with nothing. The donation games satisfies equal-gains-from-switching \citep{nowak:AAM:1990}, and hence the difference in payoff to one player due to changing strategies is independent of the partner's strategy. This convenient property often significantly simplifies the analysis.

Another well-studied but less stringent social dilemma is the snowdrift game \citep{sugden:book:1986}, which is motivated by the anecdotal story of two drivers being trapped on either side of a snowdrift with the options to either get out and start shoveling (cooperate) or to stay in the cozy warmth of the car (defect). If both drivers shovel, each gets the benefit of getting home, $b$, and they split the cost of shoveling, $0<c/2<b$. The temptation to defect exists just as in the prisoner's dilemma, but here the social dilemma is relaxed because, when facing a defector, it is now better to do all the work and get $b-c>0$ instead of nothing (and wait for spring to melt the snowdrift). That is, defection is no longer dominant. We focus on these two most prominent social dilemmas even though further and even weaker forms exist \citep{hauert:JTB:2006a}.

A generic interaction between two individuals and two strategic types, $C$ and $D$, can be characterized by a $2\times2$ payoff matrix,
\begin{align}
\label{eq:abcd}
\bordermatrix{
& C & D \cr
C & R & S \cr
D & T & P \cr} .
\end{align}
A $C$-type gets a payoff of $R$ when interacting with another $C$ and $S$ when interacting with $D$. A $D$-type gets $T$ when interacting with $C$ and $P$ when interacting with another $D$. The characteristics of the interaction are determined by the ranking of the payoffs. For example, the prisoner's dilemma is defined by $T>R>P>S$. Since $T>R$ and $P>S$, the strategy $D$ is dominant, and $R>P$ creates the social dilemma. When setting $R=b-c$, $S=-c$, $T=b$, and $P=0$, we obtain the donation game. Similarly, the snowdrift game satisfies $T>R>S>P$, with the common parametrization $R=b-\frac{1}{2}c$, $S=b-c$, $T=b$, and $P=0$ \citep{hauert:Nature:2004}. Note that for $c>b$, the snowdrift game effectively reverts to a prisoner's dilemma and defection is again the dominant strategy.

In well-mixed populations, every individual is equally likely to interact with anyone else. For a fraction, $x$, of type $C$ individuals (and $1-x$ of type $D$), the expected payoff of cooperators and defectors is $f_{C}=xR+\left(1-x\right) S$ and $f_{D}=xT +\left(1-x\right) P$, respectively. The replicator dynamics states that the type with the higher fitness increases in frequency:
\begin{align}
\dot{x} &= x\left(1-x\right)\left(f_{C}-f_{D}\right) .
\end{align}
The dynamics admit up to three equilibria: two trivial ones at ${\bf P_0}=0$ and ${\bf P_1}=1$, and possibly a third equilibrium at ${\bf Q} = \left(P-S\right)/\left(R-S-T+P\right)$, provided that ${\bf Q}\in\left(0,1\right)$. The stability of each equilibrium is easily determined by checking whether a rare type can invade. The stability criterion reduces to $P>S$ for ${\bf P_{0}}$ and to $R>T$ for ${\bf P_1}$. The equilibrium $\mathbf{Q}$ exists if both trivial equilibria are either stable or unstable, and, in the latter case, $\mathbf{Q}$ is stable.

For the donation game, the equal-gains-from-switching property (additivity) reduces the replicator dynamics to $\dot x=x\left(1-x\right)\left(-c\right)<0$. It follows that the fraction of cooperators is always dwindling, and ${\bf P_0}$ is the only stable equilibrium. In contrast, for the snowdrift game, we obtain a stable interior equilibrium at ${\bf Q}=\left(b-c\right)/\left(b-c/2\right)$, which means that cooperators and defectors can coexist as a consequence of the relaxed social dilemma.

\subsection{Asymmetric evolutionary games}
Asymmetric interactions can arise if individuals have differential access to environmental resources. More specifically, we consider a population where each individual resides on its own patch. Patches can be rich or poor in quality.

\subsubsection{Additive environmental benefits}
Rich patches confer an environmental benefit, $e$, to its occupant, whereas poor patches confer nothing. Naturally, this heterogeneity in patch quality results in asymmetric interactions that are captured by the extended, generic payoff matrix,
\begin{align}
\label{eq:abcde}
\bordermatrix{
& C_p & D_p \cr
C_r & R+e,\ R & S+e,\ T \cr
D_r & T+e,\ S & P+e,\ P \cr},
\end{align}
where $C_i$, $D_i$ with $i\in\{r,p\}$ refer to the strategic types on patches of quality $i$. The first entry in each cell of the matrix indicates the payoffs to the row player (on a rich patch), while the second entry denotes the payoffs to the column player (on a poor patch). For interactions between individuals on patches of identical quality, the payoff matrix remains symmetric and equal to
\begin{subequations}
\begin{align}
\label{eq:abcderp}
\bordermatrix{
& C_r & D_r \cr
C_r & R+e & S+e \cr
D_r & T+e & P+e \cr} ;
\qquad &\textrm{(both patches rich)} \\ 
\bordermatrix{
& C_p & D_p \cr
C_p & R & S \cr
D_p & T & P \cr} .
\qquad &\textrm{(both patches poor)}
\end{align}
\end{subequations}
The environmental benefit corresponds to an increase in the background fitness of its occupant.

The asymmetry of interactions results in four types: cooperators and defectors on good and bad patches, respectively. Thus, the state of the population, $\bf x$, is determined by four dynamical variables: ${\bf x}=(x_r, x_p, y_r, y_p)$ where $x_i$ refers to the frequency of cooperators and $y_i$ to that of defectors on rich and poor patches, respectively. Formally, instead of dealing with \emph{asymmetric} interactions of two strategies in two environments, the payoffs for the four competing types can be represented by a \emph{symmetric} $4\times4$ payoff matrix ${\bf A} = \left[a_{ij}\right]$ where $a_{ij}$ denotes the payoff of an individual of type $i$ against one of type $j$ with $i,j\in\{C_r, C_p, D_r, D_p\}$. The average payoff for each type is then given by a vector, ${\bf \pi} = {\bf A}\cdot{\bf x}$. That is, $\pi_{1}$ and $\pi_{2}$ (resp. $\pi_{3}$ and $\pi_{4}$) represent the payoffs to cooperators (resp. defectors) on rich and poor patches, respectively.

The dynamics of the four types are then governed by the system of equations, 
\begin{subequations}
\label{eq:dynnoalpha}
\begin{align}
\dot x_{r} &= f_{C} y_{r} - f_{D} x_{r} ; \\
\dot x_{p} &= f_{C} y_{p} - f_{D} x_{p} ; \\
\dot y_{r} &= f_{D} x_{r} - f_{C} y_{r} ; \\
\dot y_{p} &= f_{D} x_{p} - f_{C} y_{p} ,
\end{align}
\end{subequations}
where $f_{C}$ and $f_{D}$ denote the average fitness of cooperators and defectors, respectively. The terms $f_i x_j$ (or $f_i y_j$) indicate the (average) rates at which individuals of type $i$ produce offspring weighted by the probability that it replaces a cooperator (defector) on a patch of type $j$. For example, the frequency of $x_r$ increases whenever $C$ reproduces (which happens at a rate $f_C$) \emph{and} its offspring replaces a defector on a rich patch (which occurs with probability $y_r$) but decreases if $D$ reproduces (at rate $f_D$) \emph{and} replaces a cooperator on a rich patch (with probability $x_r$).

Naturally, the sum of the frequencies of all four types must remain normalized and add up to one, $x_r+x_p+y_r+y_p=1$. The frequency of cooperators is $x=x_r+x_p$ and the proportion of rich patches remains constant, $x_r+y_r=\alpha$ for some fixed $\alpha$ because patch qualities do not change. In the absence of correlations between strategies and patches, we have $x_r=x\alpha$, $x_p=x\left(1-\alpha\right)$, $y_r=\left(1-x\right)\alpha$, $y_p=\left(1-x\right)\left(1-\alpha\right)$, and it follows that $x$ alone still captures the full dynamics. In fact, even if initial configurations are chosen with strong correlations between strategies and patches, these correlations are eventually eliminated by the dynamics (see \app{dynnoalpha}). 

\begin{figure}
\centering
\includegraphics[width=1\textwidth]{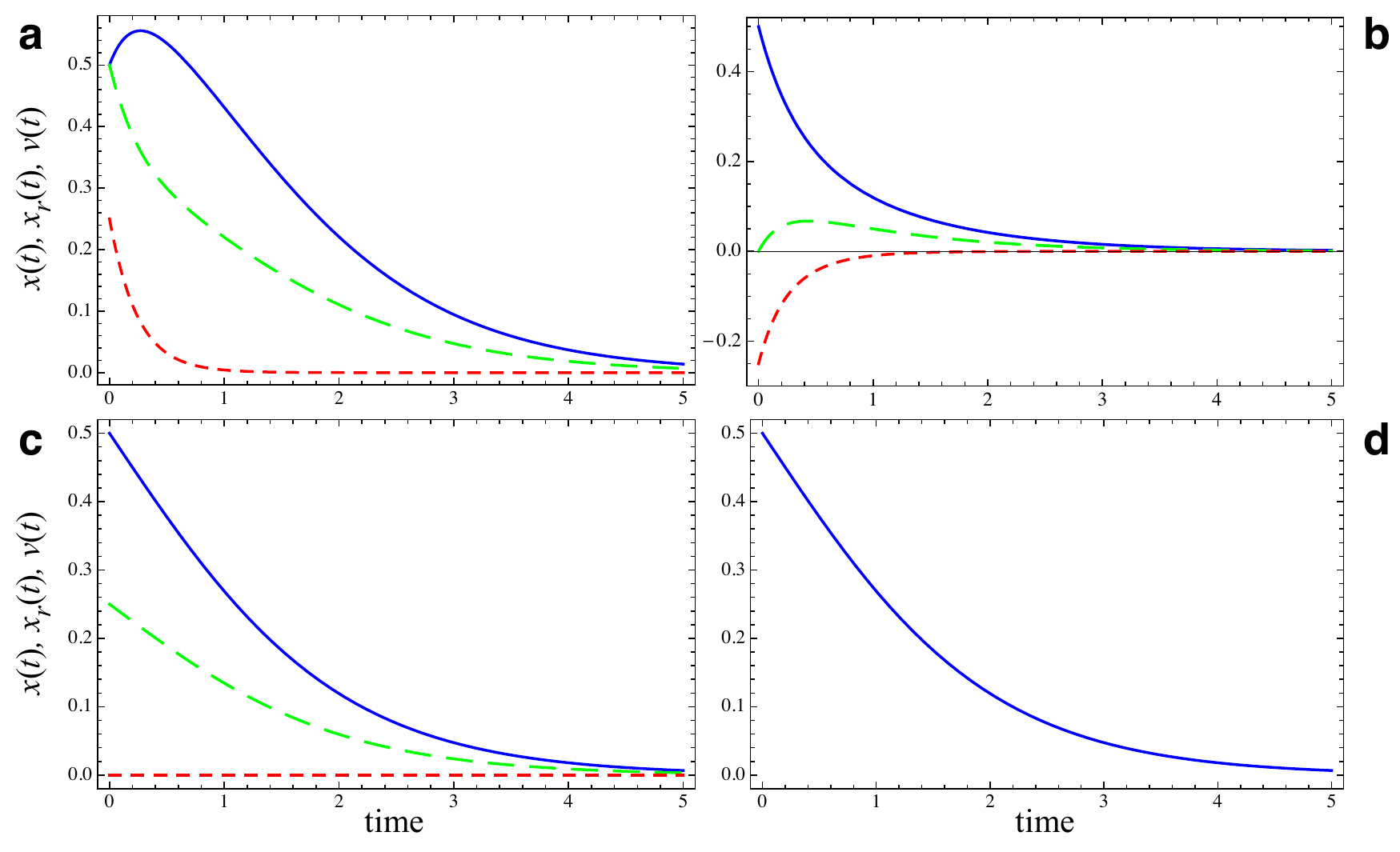}
\caption{\label{fig:abcde}
\footnotesize\sffamily 
Asymmetric donation game. Evolutionary trajectory of the frequency of cooperators ($x=x_r+x_p$, solid blue line), the frequency of cooperators on rich patches ($x_r$, long-dashed green line), and the covariance (or linkage) between cooperators and rich patches ($\nu=x_r-(x_r+x_p)(x_r+y_r)$, short-dashed red line) in asymmetric donation games with $b=4$, $c=1$, and $50\%$ rich patches ($\alpha=x_r+y_r=0.5$) that provide environmental benefits $e=2$ to their occupant. The dynamics readily eliminates any covariance introduced by initial configurations and cooperators invariably dwindle and eventually disappear: \textbf{\sffamily a} positive covariance for an initial configuration with $50\%$ cooperators and $50\%$ defectors, with all cooperators on rich patches and all defectors on poor patches; \textbf{\sffamily b} negative covariance with all cooperators on poor patches and all defectors on rich patches; \textbf{\sffamily c} initial configuration without linkage: cooperators and defectors each split equally between rich and poor patches; \textbf{\sffamily d} frequency of cooperators in the symmetric, averaged donation game in the absence of environmental benefits.
}
\end{figure}

This interpretation of \eq{dynnoalpha} implies that fitness must be non-negative, $f_C, f_D\geqslant 0$, because they represent rates of reproduction. Conversely, the average payoffs $\pi_i$ may be negative for some $a_{ij}$, for example in (\textit{i}) the donation game or (\textit{ii}) the snowdrift game in case costs exceed benefits (which essentially turns the interaction into a prisoner's dilemma). Fortunately, negative payoffs can be easily resolved by appropriately mapping payoffs to fitness. More specifically, we introduce a constant background fitness, $\sigma\geqslant -\min_{i,j} a_{ij}$, so that the rates of reproduction averaged across patches for cooperators and defectors, respectively, are
\begin{subequations}
\label{eq:fcfd}
\begin{align}
f_{C} &= x\left(\sigma+\alpha \pi_{1}+\left(1-\alpha\right)\pi_{2}\right) ; \\
f_{D} &= \left(1-x\right)\left(\sigma+\alpha \pi_{3}+\left(1-\alpha\right)\pi_{4}\right) .
\end{align}
\end{subequations}
Interestingly, the fixed points of the dynamics defined by Eq.~(\ref{eq:dynnoalpha}) and the asymmetric payoffs Eqs.~(\ref{eq:abcde})-(\ref{eq:abcderp}) are the same as those of the replicator equation with the \emph{symmetric} $2\times2$ payoff matrix
\begin{align}
\label{eq:symabcde}
\bordermatrix{
& C & D \cr
C & R+\alpha e & S+\alpha e \cr
D & T+\alpha e & P+\alpha e \cr} ,
\end{align}
which, notably, turns out to be independent of $\sigma$ (for details see \app{dynnoalpha}). Moreover, since the replicator dynamics are invariant under adding a constant to any (or all) columns of the payoff matrix, the replicator dynamics for matrix~(\ref{eq:symabcde}) are actually identical to those of matrix~(\ref{eq:abcd}), see \fig{abcde}. The results can be verified and the dynamics further explored through interactive online tutorials \citep{hauert:EvoLudo:2018}. Thus, in this scenario, environmental asymmetries may introduce only small and fleeting differences in the transient dynamics while leaving the long-term results unchanged.

\subsubsection{Asymmetric costs and benefits}
Naturally, many kinds of environmentally-induced asymmetries exist. Another kind of asymmetry is easily motivated in the donation game where costs incurred and benefits provided depend on the patch quality. More specifically, cooperators on rich patches provide benefits, $b_r$, at a cost, $c_r<b_r$, whereas on poor patches the benefit is $b_p$ and the cost is $c_p<b_p$. Defectors neither provide benefits nor incur costs. The corresponding payoff matrix is
\begin{align}
\label{eq:asymDG4x4}
\hspace{-3ex}{\bf A} = \bordermatrix{
& C_r & C_p & D_r & D_p \cr
C_r & b_r-c_r & b_p-c_r & -c_r & -c_r \cr
C_p & b_r-c_p & b_p-c_p & -c_p & -c_p \cr
D_r & b_r & b_p & 0 & 0 \cr
D_p & b_r & b_p & 0 & 0 }.
\end{align}
For $c_p>0>c_r$ this setup mimics an intriguing situation where cooperators exploit rich patches and provide benefits to their interaction partners at no cost, or more precisely, even at a personal gain of $-c_r$.

The average rates of reproduction for cooperators and defectors in the absence of correlations between strategies and patches amounts to
\begin{subequations}
\begin{align}
f_C =& x\left(x\bar b-\bar c\right) ; \\
f_D =& \left(1-x\right) x\bar c ,
\end{align}
\end{subequations}
where $\bar b=\alpha b_r+(1-\alpha)b_p$ and $\bar c=\alpha c_r+(1-\alpha)c_p$ denote the benefits and costs averaged across patch types. Thus, the change in cooperator frequencies is independent of $b_r, b_p$ and an increase requires $\bar c<0$, which effectively turns the donation game into a harmony game. However, this can only hold if cooperators exploit rich patches, $c_r<0$, and those patches are sufficiently abundant.

Similarly, asymmetries in the snowdrift game arise when drivers encounter a snowdrift on a steep mountain road. While the benefit, $b$, remains unaffected and, for simplicity, is assumed to be the same for both individuals, the costs for clearing the snow are smaller for the individual on the mountain side, $c_m$, than on the valley side, $c_v$. These cost differences could be caused, for example, by gravity, which reduces the effort needed to clear the snow. However, we note that the interpretation is not especially important here because the snowdrift game is not intended to model actual drivers on a mountain road; rather, this game simply captures important qualitative features of a relaxed social dilemma \citep{hauert:Nature:2004}, and we focus on cost heterogeneity to understand how asymmetry affects the evolutionary dynamics of this well-studied social dilemma.

With heterogeneity in costs, it is no longer obvious how to split the cost of clearing a snowdrift among a pair of cooperators. Two natural approaches are that either (\textit{i}) each clears half of the snowdrift at costs of $c_m/2$ and $c_v/2$, respectively, resulting in the payoff matrix
\begin{align}
\label{eq:asymsd}
\bordermatrix{
& C_v & D_v \cr
C_m & b-\frac{c_{m}}{2},\ b-\frac{c_{v}}{2} & b-c_{m},\ b \cr
D_m & b,\ b-c_{v} & 0,\ 0 \cr} ,
\end{align}
or (\textit{ii}) each pays the same effective cost, which results in a fair split, giving the payoff matrix
\begin{align}
\label{eq:fairsd}
\bordermatrix{
& C_v & D_v \cr
C_m & b-\tilde c,\ b-\tilde c & b-c_{m},\ b \cr
D_m & b,\ b-c_{v} & 0,\ 0 \cr},
\end{align}
where $\tilde{c}=c_{m}c_{v}/\left(c_{m}+c_{v}\right)$ \citep{du:EL:2009}. We note that costs can be split in other ways \citep{mcavoy:PLoSCB:2015} but these two cases represent particularly interesting and relevant scenarios to derive asymmetric variants of the classical snowdrift game. Naturally, snowdrifts can also arise at peaks and troughs, in which case players on both sides experience the same costs. These interactions follow a symmetric matrix,
\begin{align}
\label{eq:mvsd}
\bordermatrix{
& C_i & D_i \cr
C_i & b-\frac{c_i}2 & b-c_i \cr
D_i & b & 0 \cr},
\end{align}
where $i\in\{m,v\}$. As before, large costs ($c_i>b$) can result in negative payoffs, but including a background fitness, $\sigma$, with $\sigma>c_v$ prevents negative reproduction rates. This background fitness, $\sigma$, does not affect the population dynamics (see \app{dynnoalpha}).

\begin{figure}
\centering
\includegraphics[width=1\textwidth]{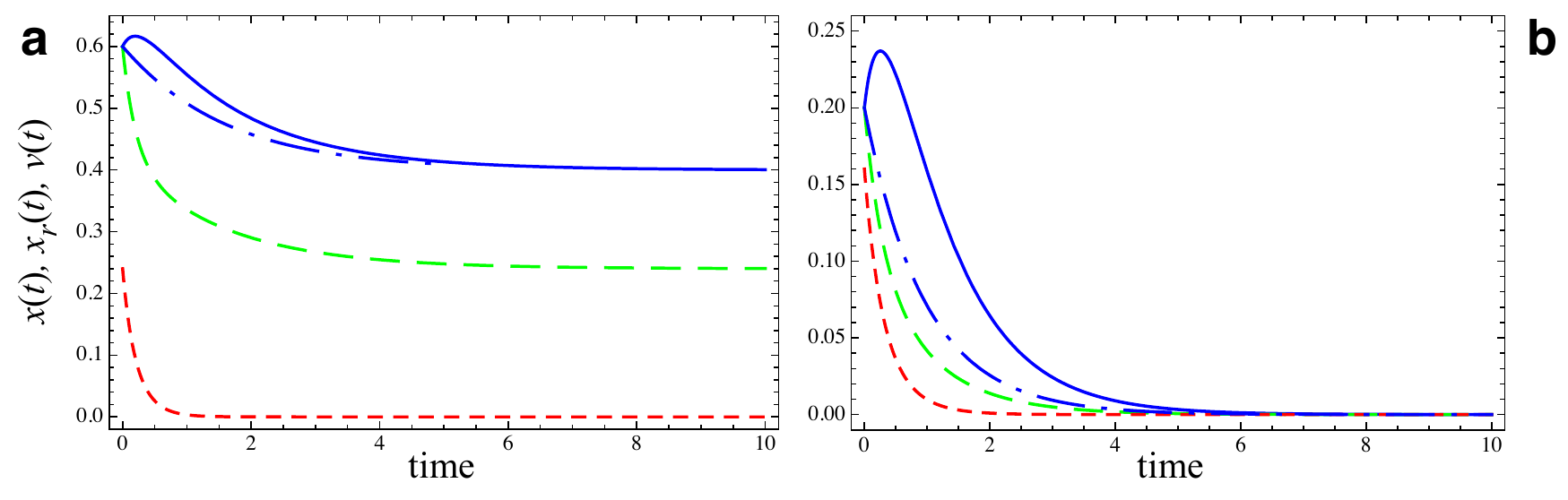}
\caption{\label{fig:sdequal}
\footnotesize\sffamily 
Asymmetric snowdrift game. Evolutionary trajectories of the frequency of cooperators ($x=x_r+x_p$, solid blue line), the frequency of cooperators on rich patches ($x_r$, long-dashed green line), and the covariance (or linkage) between cooperators and rich patches ($\nu=x_r-(x_r+x_p)(x_r+y_r)$, short-dashed red line) in asymmetric snowdrift games with equal benefits, $b=4$, but low costs on rich patches (on the mountain side), $c_m=1$, and high costs on poor patches (on the valley side), $c_v=6$. Since $c_v>b$, the snowdrift game effectively turns into a prisoner's dilemma for players on the valley side. In order to prevent negative reproduction rates, the background fitness is set to $\sigma=2$. The evolutionary outcome depends on the fraction of the population, $\alpha$, residing on rich, mountain-side patches. \textbf{\sffamily a} with sufficiently many rich patches, $\alpha=0.6$, the population, on average, is engaging in a symmetric snowdrift game with $\overline{c}=\alpha c_m+(1-\alpha)c_v=3$, which results in an equilibrium fraction of cooperators of $x^\ast=0.4$ with $x^\ast\alpha=6/25$ on rich patches. The initial configuration exhibits strong linkage with $60\%$ cooperators, all on rich patches, which results in transient differences from the frequency of cooperators obtained for the averaged game (blue, dash-dotted line). \textbf{\sffamily b} with fewer rich patches, $\alpha=0.1$, the population, on average, is essentially trapped in a prisoner's dilemma because $b<\overline{c}=5$. Consequently, the favourable conditions on rich patches remain insufficient to rescue cooperation from extinction.
}
\end{figure}

For equal (as opposed to fair) cost divisions, \eq{asymsd}, the dynamics in the long run are governed by the symmetric payoff matrix
\begin{align}
\label{eq:symsd}
\bordermatrix{
& C & D \cr
C & b-\frac{\overline{c}}2 & b-\overline{c} \cr
D & b & 0 \cr},
\end{align}
where $\overline{c}=\alpha c_m+\left(1-\alpha\right) c_v$ denotes the weighted average of costs across patches, assuming that a fraction $\alpha\ \left(=x_r+y_r\right)$ lie on the mountain side.
In contrast to the interactions in \eq{symabcde}, the proportion of rich and poor patches now does affect the population dynamics and can even change the characteristic features of the interaction from the perspective of the population. For example, if $c_m<b$ but $c_v>b$ then individuals on the mountain side interact in a snowdrift game whereas those on the valley side face a prisoner's dilemma. Depending on the abundance of rich, mountain-side patches, the population as a whole might be trapped in a prisoner's dilemma, 
and cooperators would ultimately disappear. However, when rich patches are sufficiently abundant, cooperators are able to coexist with defectors (see \fig{sdequal}).

\begin{figure}[tp]
\centering
\includegraphics[width=1\textwidth]{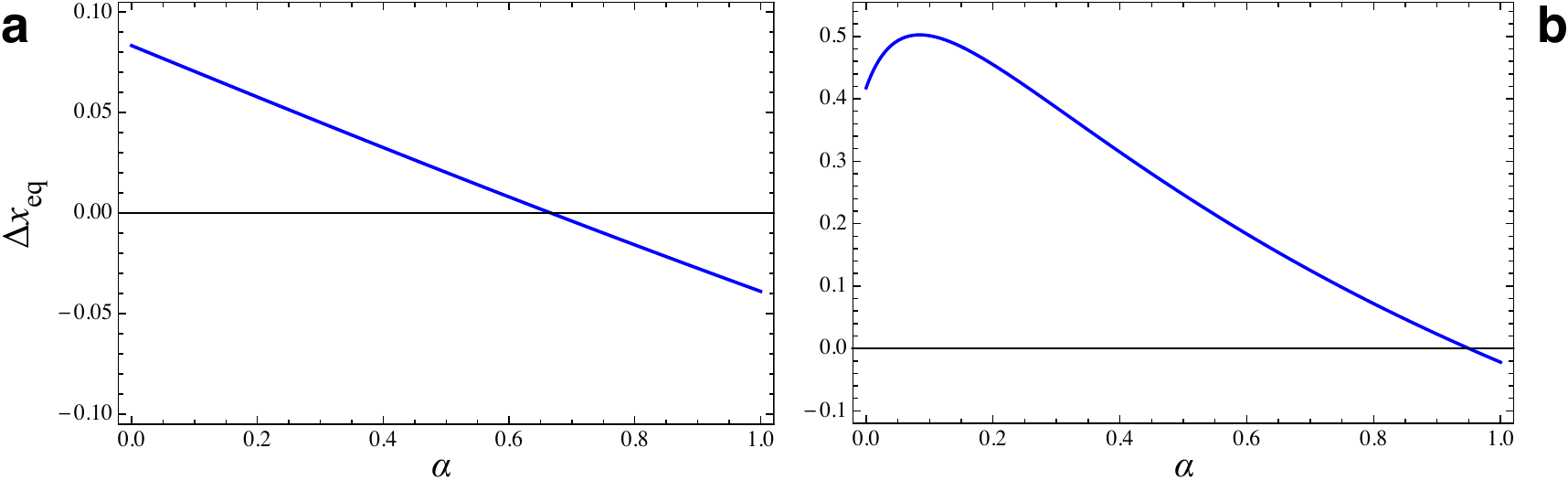}
\caption{\label{fig:sdfairvsequal}
\footnotesize\sffamily 
Fair versus equal splits of the costs in the asymmetric snowdrift game, \eq{symsd} versus \eq{symfairsd}. The difference in equilibrium fractions of cooperators for the two scenarios, $\Delta x_\text{eq}=x^\ast_\text{fair}-x^\ast_\text{equal}$, is shown as a function of the fraction of rich, mountain sites, $\alpha$. \textbf{\sffamily a} The benefit is $b=4$ and the costs are fairly similar, $c_m=1$, $c_v=2$. For $\alpha<2/3$ fair splits of the costs increase cooperation, whereas for $\alpha>2/3$ equal splits result in more cooperation. \textbf{\sffamily b} For larger differences in the costs, $c_m=0.2$, $c_v=3.8$, the effect of fair splits is more pronounced with up to $50\%$ more cooperators. Interestingly, the effect is strongest at an intermediate value, $\alpha_\text{max}\approx0.084$. Only for $\alpha>0.95$ do fair splits have a slightly detrimental impact on cooperation. 
}
\end{figure}

Similarly, for fair (but not necessarily equal) cost divisions, \eq{fairsd}, the long-term dynamics are governed by the symmetric matrix
\begin{align}
\label{eq:symfairsd}
\bordermatrix{
& C & D \cr
C & b-\tilde c & b-\overline{c} \cr
D & b & 0 \cr} .
\end{align}
Notably, only the payoff of cooperators against defectors depends on the fraction of rich sites, $\alpha$ (through $\overline{c}$). Interestingly, fair splits of the work do not necessarily support cooperators (see \fig{sdfairvsequal}).
More precisely, if $b>c_v>c_m$, i.e. all interactions are true snowdrift games, then fair splits of the work indeed promote cooperation, provided that
\begin{align}
\label{eq:sdfairfavorcoop}
\alpha<\frac{c_v}{c_v+c_m}.
\end{align}
Otherwise, equal splits of the work result in larger fractions of cooperators at equilibrium. In fact, the scale tips in favor of fairness if the expected costs of cooperation are higher in the valley than on the mountain, $c_v(1-\alpha)>\alpha c_m$. For $\alpha<0.5$, i.e. if less than half the population resides on rich, mountain-side sites, fair splits always results in increased cooperation. Only for sufficiently large $\alpha$ is it the case that fair splits no longer provide an advantage to cooperators.

\section{Environmental feedback -- the conservation game}
In the above analysis, individuals interact with each other, but they do not change their environment. As a result, the fraction of rich and poor patches remains constant. Let us now turn to the other extreme, where individuals interact exclusively with their environment. As before, a rich patch provides a benefit, $e$, to the occupant, whereas a poor patch provides nothing. However, cooperators now tend to their patch at a cost, $c$, which maintains the quality of rich patches and restores poor patches at a fixed rate, $\lambda$. Conversely, defectors exploit their patches so that poor patches are unable to recover, and rich patches deteriorate and turn into poor ones at rate $\lambda\rho$. With this parametrization, $\rho$ indicates how much faster (or slower) patch degradation occurs as compared to restoration, while the common factor $\lambda$ determines the timescales between evolutionary and ecological processes: for large $\lambda$ (relative to the benefit, $e$, and the cost, $c$) ecological processes happen faster than evolutionary changes (and vice versa for small $\lambda$). Thus, cooperators and defectors engage in a game of environmental conservation.

The state of the population is again captured by ${\bf x}=(x_r, x_p, y_r, y_p)$ where $x_i$ ($y_i$) denote the frequency of cooperators (defectors) on rich and poor patches. The dynamics are defined by
\begin{subequations}
\label{eq:dynalpha}
\begin{align}
\dot x_r &= f_C y_r - f_D x_r + \lambda x_p ; \\
\dot x_p &= f_C y_p - f_D x_p - \lambda x_p ; \\
\dot y_r &= f_D x_r - f_C y_r - \lambda\rho y_r ; \\
\dot y_p &= f_D x_p - f_C y_p + \lambda\rho y_r .
\end{align}
\end{subequations}
For $\lambda=0$, these equations reduce to the previous dynamics, \eq{dynnoalpha}, without environmental feedback and with constant patch qualities. The fitness or, more precisely, the reproductive outputs of cooperators and defectors, respectively, averaged across patch types are
\begin{subequations}
\begin{align}
f_C &= x\left(\sigma-c\right)+x_r e ; \\
f_D &= \left(1-x\right)\sigma+y_r e ,
\end{align}
\end{subequations}
where $x=x_r+x_p$ and $\sigma$ again denotes a background fitness with $\sigma\geqslant c$ to ensure non-negative rates of reproduction. The constraint $x_r+x_p+y_r+y_p=1$ must still hold, but, in contrast to \eq{dynnoalpha}, no other simplifications are possible, which results in three independent dynamical variables.

A different but more intuitive and illustrative set of dynamical variables is given by $x=x_r+x_p$, which tracks evolutionary changes in the frequency of cooperators; $\alpha=x_r+y_r$, which tracks ecological changes in the abundance of rich patches; and the covariance between cooperators and rich patches, $\nu=x_r-x\alpha$, which indicates the linkage beyond random assignments of individuals to patches (see \app{transform}). The corresponding dynamical equations of the conservation game are
\begin{subequations}
\begin{align}
\dot x &= \left(1-x\right)x\left(-c\right)+\nu e ; \\
\dot \alpha &= \lambda\left( x \left(1+\alpha\left(\rho-1\right)\right) + \nu\left(\rho-1\right) - \alpha\rho \right) ; \\
\dot \nu &= x \left( \left(1-x\right)\lambda\left(1+\alpha\left(\rho-1\right)\right) + \nu\left(c-\lambda\left(\rho-1\right)\right) \right) - \nu\left(e\alpha+\sigma+\lambda\right) .
\end{align}
\end{subequations}
The dynamics admit up to three fixed points of the form $\left(\hat x, \hat\alpha, \hat\nu\right)$. Two trivial fixed points, ${\bf P_0}=\left(0,0,0\right)$ and ${\bf P_1}=\left(1,1,0\right)$, always exist and represent homogeneous states with all defectors on poor patches ($\mathbf{P_{0}}$) and all cooperators on rich patches ($\mathbf{P_{1}}$), respectively. If it exists, the third, interior equilibrium $\textbf{Q}$ satisfies
\begin{subequations}
\label{eq:q}
\begin{align}
\hat x &= \frac\rho c\frac{\lambda(e-c)-c\sigma}{e-c-(\rho-1)\sigma} ; \\
\hat\alpha &= \frac{\hat x+(\rho-1)\hat\nu}{\rho-(\rho-1)\hat x} ; \\
\hat\nu &= \frac c e \hat x(1-\hat x) ,
\end{align}
\end{subequations}
which indicates a mixture of cooperators and defectors on rich and poor patches. Clearly, smaller maintenance costs, $c$, and larger environmental benefits, $e$, both support cooperation, i.e. increase $\hat x$; see \eq{q}. Moreover, cooperation is supported by faster ecological changes (increasing $\lambda$) and faster degradation of rich patches by defectors (increasing $\rho$).

Stability analysis shows that the trivial equilibrium ${\bf P_0}$ is a stable node if $\left(e-c\right)\lambda<c\sigma$ (see \app{conservation}). Consequently, for sufficiently large environmental benefits, $e$, sufficiently small maintenance costs, $c$, or sufficiently fast ecological changes, $\lambda>\lambda_c=c\sigma/\left(e-c\right)$, the defector equilibrium turns into an unstable node (but never an unstable focus) and cooperators can persist. Conversely, the cooperator equilibrium ${\bf P_1}$ is a stable node if the combined ecological effects exceed a threshold, $\lambda\rho>c\left(e-c+\sigma\right)/\left(e-c\right)$. Therefore, the homogeneous state consisting of only cooperators is always stable for sufficiently small costs, $c$, or sufficiently fast ecological changes. Note that ${\bf P_1}$ might be a focus, but in this case, it is stable only in the unfortunate situation where patches require maintenance costs that exceed their environmental benefit, $c>e$ (see \app{conservation}).

Unfortunately, the stability analysis of the interior equilibrium involves unwieldy solutions to cubic polynomials and hence remains inaccessible in closed form. Nevertheless, some inferences can be made based on the stability of ${\bf P_0}$ and ${\bf P_1}$. If both homogeneous equilibria are unstable, we can expect ${\bf Q}$ to exist and be either stable (node or focus) or unstable (focus). In the latter case, the dynamics should undergo a Hopf bifurcation and exhibit a stable limit cycle.

\subsection{Ecological changes}
\begin{figure}
\centering
\includegraphics[width=0.6\textwidth]{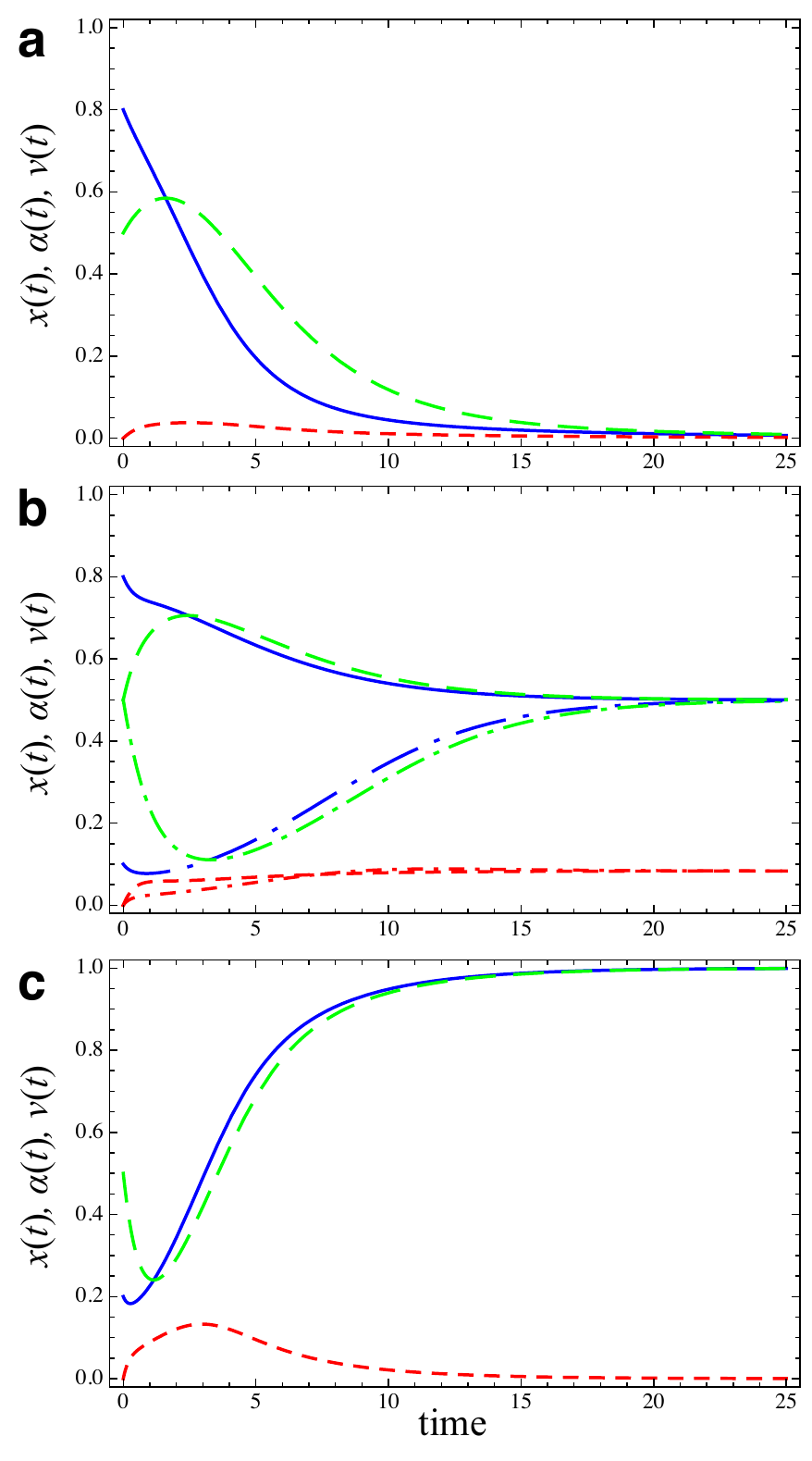}
\caption{\label{fig:lambda}
\footnotesize\sffamily 
Effects of ecological versus evolutionary changes in the conservation game, where rich patches provide an environmental benefit, $e=3$, cooperators pay costs, $c=1$, to maintain rich patches or to restore poor ones, and the background fitness, $\sigma =1$, prevents negative rates of reproduction. For equal degradation and restoration rates, $\rho=1$, the long-term behaviour is determined by the ecological timescale, $\lambda$. Panels \textbf{\sffamily a}-\textbf{\sffamily c} illustrate the increase in cooperation (solid blue line), rich patches (long-dashed green line) and the covariance between the two (short-dashed red line) for increasing rates of ecological changes, larger $\lambda$. %
\textbf{\sffamily a} for slower changes ($\lambda=0.4$), the environment deteriorates and defection dominates. %
\textbf{\sffamily b} for intermediate rates ($\lambda=1$), cooperators and defectors coexist regardless of the initial configuration (the solid/dashed and dash-dotted lines give initial cooperator fractions of $x_0=0.8$ and $x_0=0.1$, respectively).  %
\textbf{\sffamily c} for faster changes ($\lambda=2$), cooperators efficiently generate rich patches and readily take over.}%
\end{figure}

The rate of ecological changes as compared to evolutionary changes is controlled by the parameter $\lambda$, whereas the kind of ecological changes depends on the rate of degradation as compared to restoration, $\rho$. In order to highlight the effects of ecological changes, we consider the special case of equal degradation and restoration rates, $\rho=1$. In this case, the analysis becomes simpler and, in particular, the interior fixed point is now given by $\textbf{Q}=\left(\hat x, \hat x, \left(1-\hat x\right)\hat x c/e\right)$ with $\hat x=\lambda/c-\sigma/\left(e-c\right)$. Interestingly, at $\textbf{Q}$, cooperators and rich patches have the same frequency, which tends to increase for lower maintenance costs, $c$, smaller environmental benefits, $e$, or faster ecological changes, $\lambda$, as shown in \fig{lambda}.
Stable (limit) cycles cannot occur based on the analysis of the characteristic polynomial, which shows that a Hopf bifurcation is impossible for $e>c$; see \app{hopf}.

\subsection{Restoration versus degradation}
\begin{figure}
\centering
\includegraphics[width=1\textwidth]{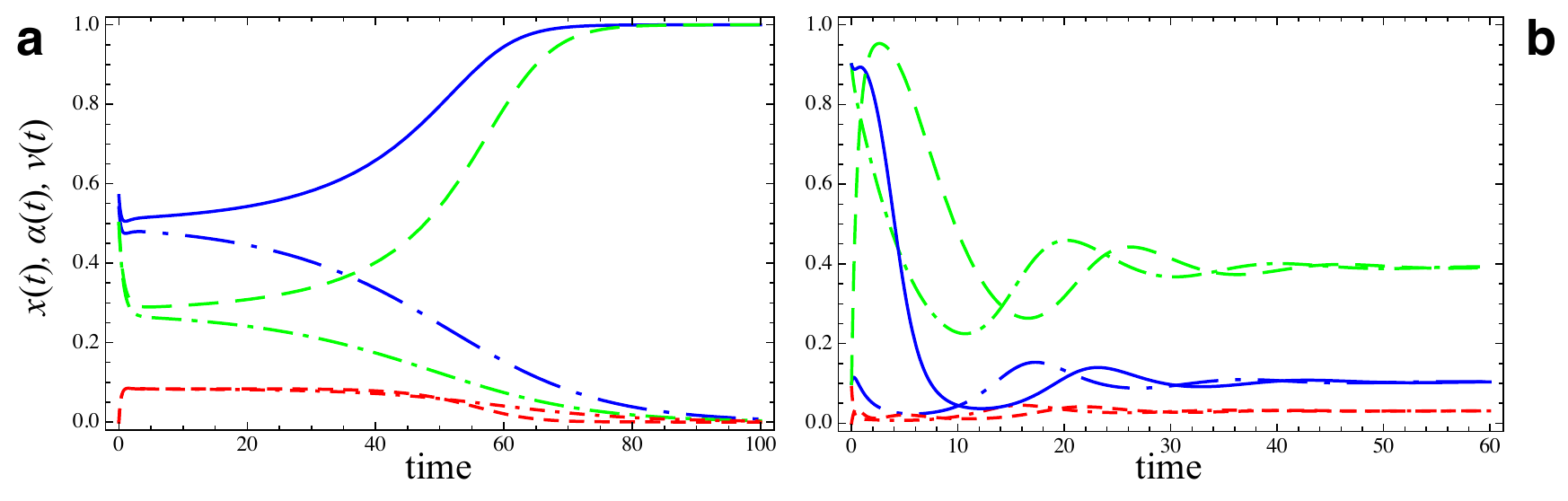}
\caption{\label{fig:rho}
\footnotesize\sffamily 
Effects of degradation versus restoration in conservation games on the frequency of cooperation (solid blue line), rich patches (long-dashed green line), and the covariance between the two (short-dashed red line). Game parameters are $e=3$, $c=1$, and $\sigma=1$, as in \fig{lambda}. 
\textbf{\sffamily a} for slow ecological changes ($\lambda=0.4$, c.f. \fig{lambda}a), high degradation rates ($\rho=5$) can save cooperation by creating a bi-stable dynamics: for initial frequencies of cooperators above some threshold ($x_0=0.57$, solid and dashed lines), cooperators thrive and eventually reach fixation; whereas for initial frequencies below the threshold ($x_0=0.54$, dash-dotted lines) the conversion of poor patches remains too slow and cooperators disappear. In both cases, the dynamics result in a clear positive linkage between cooperators and rich patches during the transient phase but gets eroded sooner for smaller $x_0$, which then results in the demise of cooperation.
\textbf{\sffamily b} for fast ecological changes ($\lambda=2$, c.f. \fig{lambda}c), slow degradation rates ($\rho=0.1$) can prevent cooperation from taking over and resulting in coexistence of cooperators and defectors, regardless of the initial configuration. Sometimes the dynamics exhibit damped oscillations as shown here for large initial fractions of cooperators on few rich patches without linkage ($x_0=0.9$, $\alpha_0=0.1$, and $\nu_0=0$; solid and dashed lines) as well as strong linkage with initially many rich patches and few cooperators but all on poor patches ($x_0=0.1$, $\alpha_0=0.9$, and $\nu_0=0.09$; dash-dotted lines).
}
\end{figure}

Faster ecological changes (larger $\lambda$) support cooperation by increasing the covariance or linkage between cooperators and rich patches, $\nu$. Similarly, increasing the rate at which defectors degrade rich patches relative to cooperators restoring poor ones (larger $\rho$) also increases $\nu$ by increasing the covariance between defectors and poor patches. For $\rho>1$, restoration of poor patches is faster than degradation of rich patches (and vice versa for $\rho<1$). More specifically, under conditions where cooperators disappear, as in \fig{lambda}a, increases in $\rho$ result in bistable dynamics, where both $\bf P_0$ and $\bf P_1$ are stable ($\textbf{Q}$ is unstable) and have their respective basins of attraction such that the evolutionary outcome depends on the initial configuration; see \fig{rho}a. 
For sufficiently large initial fractions of cooperators, rich patches, and/or linkage between the two, cooperation can thrive and eventually take over the population. Conversely, under conditions where cooperators prevail, as in \fig{lambda}c, lowering $\rho$ reduces the linkage between defectors and poor patches such that cooperators are no longer able to take over. Instead, both $\bf P_0$ and $\bf P_1$ are unstable, and a stable interior fixed point $\textbf{Q}$ appears where cooperators and defectors coexist (\fig{rho}b). The conservation game can be explored interactively \citep{hauert:EvoLudo:2018} and a summary of the qualitative dynamics as a function of the ecological parameters, $\lambda$ and $\rho$ is shown in \fig{lambdarho}a.

\section{Games \& ecology}
So far, our discussion has been limited to two extreme cases: (\textit{i}) individuals interact with each other but not with their environment or (\textit{ii}) individuals interact with their environment but not each other. In the following, we now turn to the general eco-evolutionary dynamics, which combines the two types of interactions to obtain asymmetric evolutionary games in a changing environment. 

\subsection{Eco-evolutionary dynamics}
The dynamics follow \eq{dynalpha}, but the rates of reproduction now include payoffs from interactions with other members of the population; see \eq{fcfd}. Therefore, in addition to engaging in social interactions, cooperators also restore poor patches, while defectors not only attempt to take advantage of others but also exploit and degrade the environment.

\begin{figure}
\centering
\includegraphics[width=0.8\textwidth]{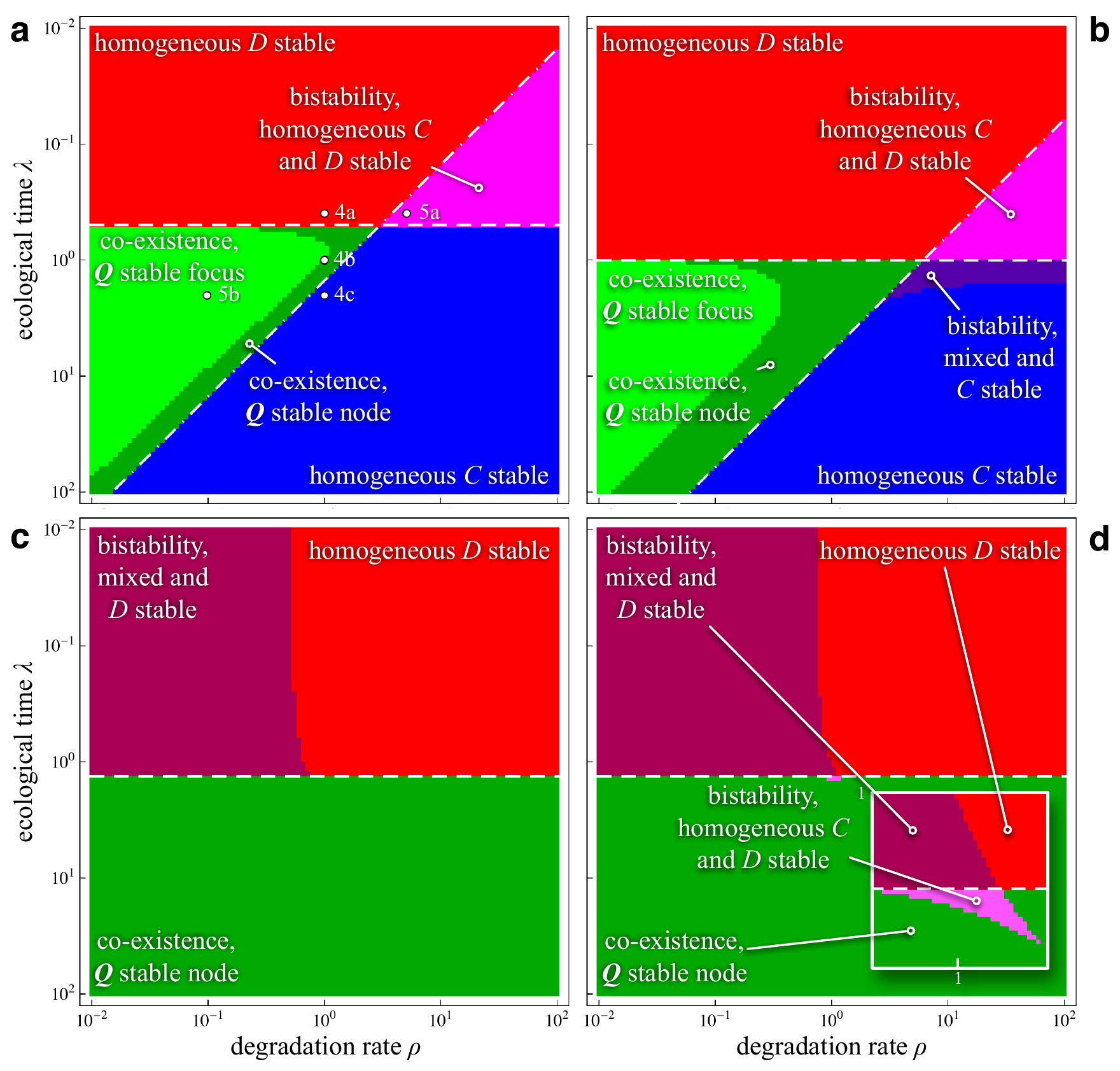}
\caption{\label{fig:lambdarho}\setstretch{1.1}
\footnotesize\sffamily 
Dynamics of evolutionary games with ecological feedback as a function of the speed of ecological changes, $\lambda$, and the relative rates of degradation of rich patches, $\rho$, by defectors as compared to restoration of poor ones by cooperators. For small $\lambda$, defection, ${\bf P_0}$, is stable (above dashed line) while for large $\rho$, cooperation, ${\bf P_1}$, is stable (below dash-dotted line).
\textbf{\sffamily a} The conservation game does not include interactions between individuals but serves as a relevant reference. The dots indicate parameter combinations depicted in \figs{lambda}{rho} with $e=3$, $c=1$, and $\sigma=1$. For small $\lambda$ and $\rho$, defection dominates and all patches are of poor quality (\emph{red}). Increasing $\rho$ strengthens linkage between cooperators and rich patches and results in bistable dynamics with defection and cooperation both stable (\emph{pink}). Similarly, increasing $\lambda$ further supports cooperation and leads to stable coexistence (\emph{green}) or even dominant cooperation (\emph{blue}).
\textbf{\sffamily b} Asymmetric donation game with environmental feedback for $e=2$, $b=4$, $c=1$, and $\sigma=1$ (c.f. \fig{abcde} without feedback). The dynamical scenarios follow the same patterns as for the conservation game in \textbf{\sffamily a} but cooperation is slightly more challenging in that larger $\rho$ or $\lambda$ are required. The only qualitative difference is a small region with two interior equilibria, which results in bistability between stable coexistence and homogeneous cooperation (\emph{purple}). 
\textbf{\sffamily c} Asymmetric snowdrift game with equal splits for $b=4$, $c_r=1$, $c_p=6$, and $\sigma=2$ (c.f. \fig{sdequal} without feedback). Sufficiently large $\lambda$ results in stable coexistence, independent of $\rho$. For smaller $\lambda$ and large $\rho$, defection dominates because the relentless restoration of poor patches becomes too costly. Decreasing $\rho$ yields two interior fixed points and results in bistability between homogeneous defection and stable coexistence (\emph{maroon}).
\textbf{\sffamily d} Same as \textbf{\sffamily c} but with fair splits of costs. The qualitative dynamical scenarios remain essentially unchanged but fair splits provide some support for cooperation by extending the region of bistability (\emph{maroon}). Moreover, a small region of bistability between homogeneous cooperator and defector states appears (\emph{pink}, see inset).
}
\end{figure}

The generic dynamics in terms of the fraction of cooperators, $x$, fraction of good patches, $\alpha$, and the covariance between the two, $\nu$, are derived in \app{transform}. Again, the two trivial fixed points, ${\bf P_0}$ and ${\bf P_1}$, always exist and correspond to the homogeneous states of defectors on poor patches and cooperators on rich patches, respectively. Unfortunately, however, the generic case cannot be analyzed in closed form for the same reasons as before and hence we present the effects of ecological feedback on the evolutionary outcomes through representative numerical investigations of the characteristic dynamics (see \fig{lambdarho} for a summary and \citet{hauert:EvoLudo:2018} for interactive tutorials).
Nevertheless, three important and illustrative special cases permit a more detailed analysis: (\textit{i}) additive environmental benefits, (\textit{ii}) equal-gains-from-switching (donation game), and, additionally, (\textit{iii}) identical restoration and degradation rates.

First, for additive environmental benefits, the rates of reproduction are given by \eq{fcfdabcde}, and the fixed points of the dynamics are
\begin{subequations}
\begin{align}
\label{eq:hataecoevo}
\hat\alpha &= \frac{\hat\nu\left(\rho-1\right)+\hat x}{\rho-\left(\rho-1\right)\hat x} ; \\
\hat\nu &= \frac{1}{e} \hat x\left(1-\hat x\right)\left(P-S-\hat x\left(R-S-T+P\right)\right) ,
\end{align}
\end{subequations}
while $\hat x$ is determined by the roots of
\begin{align}
\label{eq:hatxecoevo}
\frac{1}{e}\frac{x\left(1-x\right)}{\rho\left(1-x\right)+x}\bigg(e\lambda\rho-\left(P-S-x\left(R-S-T+P\right)\right)\times\notag\\
\Big(x\left(x R+\left(1-x\right) S+e\right)+\rho\left(1-x\right)\left(x T+\left(1-x\right) P\right)+ \notag\\
\lambda\rho+\sigma\left(\rho\left(1-x\right)+x\right)\Big)\!\bigg)=0.
\end{align}
Note that no singularities exist because the denominator in \eq{hatxecoevo} is always positive. In addition to $\mathbf{P_0}$ and $\mathbf{P_1}$, up to three additional interior fixed points may exist but, unfortunately, finding them explicitly involves solving cubic polynomials, which precludes further detailed analysis. However, just as for the conservation game (see \app{conservation}), the stability conditions at ${\bf P_0}$ and ${\bf P_1}$ remain accessible: ${\bf P_0}$ is stable for sufficiently slow ecological changes, namely $\lambda<\lambda_c=-S\sigma/\left(S+e\right)$, while ${\bf P_1}$ is stable if the combined ecological effects are sufficiently strong, namely $\lambda\rho>-\left(T-R\right)\left(R+e+\sigma\right)/\left(T-R-e\right)$. 

Second, the donation game satisfies the equal-gains-from-switching property, $R-S=T-P$, which simplifies the dynamics:
\begin{subequations}%
\begin{align}%
\dot x &= -x(1-x)c+e\nu ; \\
\dot\alpha &= \lambda\left(x-\alpha\rho +\left(\rho-1\right)\left(\alpha x+\nu\right)\right) ; \\
\dot\nu &= \lambda x \left(1-x+\left(\rho-1\right)\left(\alpha\left(1-x\right)-\nu\right)\right) \notag\\
&\qquad -\nu\left(\alpha e+\lambda+\sigma+x\left(b-c\right)\right) ,
\end{align}
\end{subequations}
with equilibria $\hat\nu=\hat x\left(1-\hat x\right) c/e$ and $\hat\alpha$ as before, \eq{hataecoevo}, and $\hat x$ given by the roots of
\begin{align}
\frac{1}{e}\frac{x\left(1-x\right)}{\rho\left(1-x\right)+x}\Big(\left(e-c\right)\left(\lambda\rho-x c\right) -c\left(\rho\left(1-x\right) +x\right)\left(\sigma+x b\right)\Big) &= 0 ,
\end{align}
which, apart from the trivial equilibria, is reduced to a quadratic polynomial admitting at most two interior equilibria. Nevertheless, the analytical expressions remain unwieldy for further analysis. A numerical classification of the dynamics are shown in \fig{lambdarho}b. For sufficiently slow ecological changes, $\lambda<\lambda_c=c\sigma/(e-c)$, defection at ${\bf P_0}$ is stable. For slow degradation rates (small $\rho$), ${\bf P_0}$ is the only stable outcome, but increasing $\rho$ strengthens linkage between cooperators and rich patches and eventually results in bistable dynamics, where the system ends up in ${\bf P_0}$ or ${\bf P_1}$ depending on the initial configuration. Faster ecological changes, $\lambda>\lambda_c$, support cooperation and result in stable coexistence for slow degradation rates, small $\rho$, and stable cooperation at ${\bf P_1}$ for $\lambda\rho>c\left(b-c+e+\sigma\right)/\left(e-c\right)$ with a small region of bistability between stable coexistence and ${\bf P_1}$, see \fig{lambdarho}b.

Third, for the donation game with equal restoration and degradation rates ($\rho =1$) and minimal background fitness ($\sigma =c$), the dynamics admit at most a single interior equilibrium, ${\bf Q}=\left(\hat x, \hat x, \hat x\left(1-\hat x\right)c/e\right)$, where $\hat x=\left(\lambda\left(e/c-1\right)+c\right)/\left(b-c+e\right)$. Interestingly, increasing benefits, $b$, tends to undermine cooperation and favours defectors. ${\bf P_0}$ becomes unstable and cooperators can persist for sufficiently fast ecological changes, $\lambda>\lambda_c=c^{2}/\left(e-c\right)$, sufficiently large environmental benefits, $e$, or sufficiently small costs of cooperation, $c$. Similarly, ${\bf P_1}$ is stable if the combined ecological effects exceed a threshold, $\lambda\rho>c\left(b-c+e+\sigma\right)/\left(e-c\right)$, which always holds for sufficiently small costs, $c$. Unfortunately, even in this restrictive setup, the stability of ${\bf Q}$ remains analytically inaccessible.

In the asymmetric snowdrift game, \eqs{asymsd}{mvsd}, ecological feedback does not affect the qualitative dynamics and coexistence is stable provided $b>c_r, c_p$. Therefore, regardless of patch quality, it always pays to adopt a strategy that is different from that of the opponent. This behavior changes if, for example, the cost on poor patches exceeds the benefits, $c_p>b>c_r$, and effectively turns those interactions into prisoner's dilemmas. In that case, stable coexistence can be maintained only for sufficiently fast ecological changes, $\lambda>\lambda_c=-\sigma\left(b-c_p\right)/\left(b-c_r\right)$, but regardless of degradation rates, $\rho$. For $\lambda<\lambda_c$ defection at ${\bf P_0}$ is stable, but slow degradation rates, $\rho$, introduce two interior fixed points and result in bistability between ${\bf P_0}$ and coexistence; see \fig{lambdarho}c. This general pattern holds regardless of whether the work for clearing the snowdrift is split equally or fairly. The latter case only extends the region of bistability; see \fig{lambdarho}d.

\subsection{Socio-environmental dilemmas}\label{subsubsec:socenv}
\begin{figure}
\centering
\includegraphics[width=0.8\textwidth]{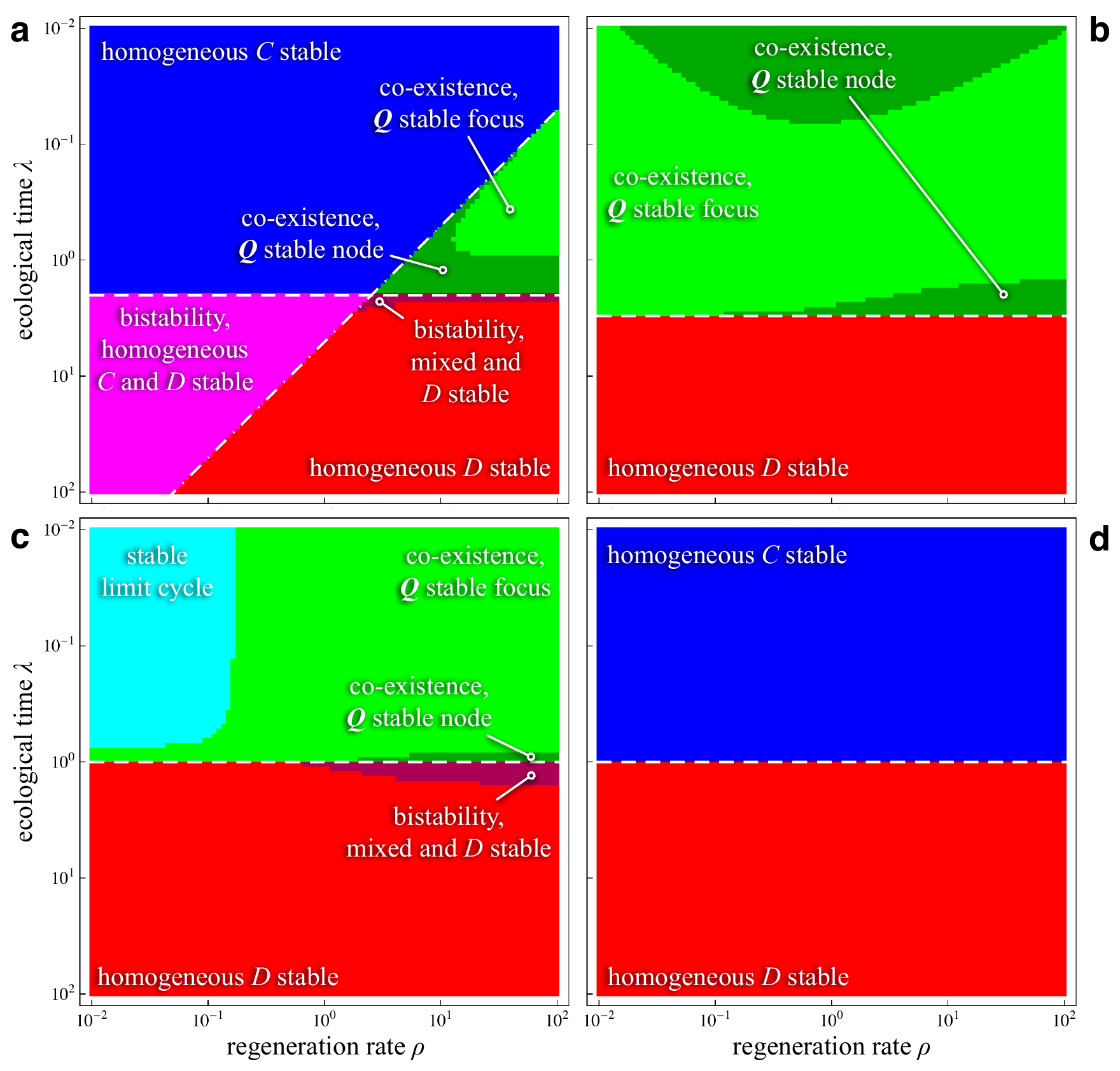}
\caption{\label{fig:lambdarhocdeg}\setstretch{1.1}
\footnotesize\sffamily 
Dynamics of evolutionary games with ecological feedback as a function of the speed of ecological changes, $\lambda$, and the relative rates of regeneration of poor patches, $\rho$, occupied by defectors as compared to degradation of rich ones through cooperators. For large $\lambda$, defection, ${\bf P_0}$, is stable (below dashed line), while for small $\rho$, cooperation, ${\bf P_1}$, is stable (above dash-dotted line).
\textbf{\sffamily a} The harmony game with additive environmental benefits for $b=4$, $c=-1$, $e=2$, and $\sigma=0$ results in a striking counterpart to the dynamical patterns of the donation game with positive impacts of cooperators on the environment, c.f. \fig{lambdarho}b. In fact, the dynamics are essentially the opposite: where cooperation was stable (\emph{blue}), defection is now stable (\emph{red}), and stable coexistence (\emph{green}) is replaced by bistability (\emph{pink}) and vice versa. Finally, a small region of bistability between defection and coexistence (\emph{maroon}) exists that exhibited bistability between cooperation and coexistence.
\textbf{\sffamily b} Asymmetric snowdrift game with equal splits of the costs for $b=4$, $c_r=1$, $c_p=6$, and $\sigma=2$ (c.f. \fig{sdequal} without feedback). For fast ecological changes (large $\lambda$), the environment is unable to recover, and defectors dominate regardless of the degradation rate, $\rho$. Conversely, small $\lambda$ results in stable coexistence of cooperators and defectors. Splitting the costs fairly produces the same qualitative dynamics (not shown).
\textbf{\sffamily c} Asymmetric donation game where cooperators exploit their patch for $b_r=b_p=4$, $c_r=-1$, $c_p=1$, and $\sigma=1$. Cooperation is dominant on rich patches but defection dominates on poor patches. For fast ecological changes, defectors again dominate -- with the exception of a small region of bistability between defection and coexistence (\emph{maroon}), which is possible due to faster regeneration rates (larger $\rho$). Cooperation is sustainable only for slow ecological changes (small $\lambda$) and results in stable coexistence (\emph{green}) or persistent oscillations through stable limit cycles (\emph{cyan}) for small $\rho$.
\textbf{\sffamily d} Same as \textbf{\sffamily c} but in the limit of opportunistic cooperators that help only when there is no cost to themselves, with $b_r=4$, $c_r=-1$, $b_p=0.01$, $c_p=0$, and $\sigma=0$. Note that the dynamics for $b_p=0$ becomes degenerate and hence the limit for small $b_p$ needs to be considered. As before, fast ecological changes cannot sustain cooperation but now cooperators dominate for small $\lambda$.
}
\end{figure}

Up to this point, we tacitly assumed that cooperators care not only about the well-being of their social partners but also care about and restore their environment. That ``cooperation" accurately describes such an action is unambiguous since both a co-player and the environment would view it as such. We now turn to a more hypocritical (or at least more pragmatic) form of cooperation in which cooperators exploit and degrade rich patches at rate $\lambda$ and prevent the regeneration of poor patches in order to provide benefits to others. Defectors, on the other hand, attempt to free ride on benefits provided by others but do not extract environmental resources, which allows poor patches to regenerate at rate $\rho\lambda$. The resulting dynamics are governed by
\begin{subequations}
\label{eq:dyncdeg}
\begin{align}
\dot{x}_{r} &= f_{C} y_{r} - f_{D} x_{r} - \lambda x_{r} ; \\
\dot{x}_{p} &= f_{C} y_{p} - f_{D} x_{p} + \lambda x_{r} ; \\
\dot{y}_{r} &= f_{D} x_{r} - f_{C} y_{r} + \lambda\rho y_{p} ; \\
\dot{y}_{p} &= f_{D} x_{p} - f_{C} y_{p} - \lambda\rho y_{p} ,
\end{align}
\end{subequations}
which is similar to \eq{dynalpha} except for the reversed ecological impacts. Thus, the analysis remains analogous to the previous scenario but analytical results remain equally inaccessible. The dynamics again admit two trivial fixed points, ${\bf P_0}=\left(0,1,0\right)$ and ${\bf P_1}=\left(1,0,0\right)$, but now ${\bf P_0}$ refers to homogeneous states of defectors on rich patches ($\hat\alpha =1$), whereas ${\bf P_1}$ indicates homogeneous cooperation on poor patches ($\hat\alpha =0$). In addition, up to three interior equilibria may again exist. Instead of focusing on generic cases, we discuss the effects of ecological feedback for a selection of relevant and illustrative types of interactions through numerical investigations of the characteristic dynamics; see \fig{lambdarhocdeg} (and \citet{hauert:EvoLudo:2018} for interactive tutorials).

The donation game with negative ``costs'' transforms the interaction into the harmony game, which renders cooperation the dominant strategy and hence eliminates the social dilemma. Including additive environmental benefits, $e$, does not change the outcome for positive feedback between cooperators and rich patches, as in \eq{dynalpha}, and cooperation (on rich patches) remains stable. However, much richer and more interesting dynamics unfold when introducing negative feedback between cooperators and rich patches. In the present context, negative costs could represent benefits extracted from the environment. Interestingly, for $e>-c>0$, i.e. if defectors on rich patches are better off than cooperators on poor ones, the resulting dynamics are essentially the opposite of those in the donation game with positive costs and positive feedback, c.f. \fig{lambdarho}b and \fig{lambdarhocdeg}a. Apart from a small shift in terms of $\lambda$ and $\rho$, the qualitative dynamics of the two scenarios simply have the stability of all fixed points inverted. For fast ecological changes, $\lambda>\lambda_c=-c\left(e+\sigma\right)/\left(c+e\right)$, cooperators efficiently degrade the environment and render homogeneous defection (on rich patches) stable at ${\bf P_0}$. However, slow regeneration rates, $\lambda\rho<-c\left(b-c+\sigma\right)/\left(c+e\right)$, increase linkage between cooperators and poor patches and render ${\bf P_1}$ stable as well, resulting in bistability. For slow ecological changes, $\lambda<\lambda_c$, the environment is unable to recover, and ${\bf P_1}$ is the sole stable outcome except for fast regeneration rates (large $\rho$), which increase linkage between defectors and rich patches and hence promote stable coexistence.

Introducing negative feedback between cooperators and rich patches for the asymmetric snowdrift game, \eqs{asymsd}{mvsd}, with equal splits of the costs results in stable coexistence for slow ecological changes, $\lambda<\lambda_c=-\sigma\left(b-c_r\right) /\left(b-c_p\right)$, but drives cooperators to extinction for fast changes, i.e. ${\bf P_0}$ is stable for $\lambda>\lambda_c$; see \fig{lambdarhocdeg}b. As long as $c_p>0$, cooperation at ${\bf P_1}$ is never stable. The qualitative dynamics are largely independent of the regeneration rate, $\rho$, and also remain essentially unaffected by
equal or fair splits of the costs.

Finally, we return to the asymmetric donation game \eq{asymDG4x4} in which cooperators on rich patches provide benefits, $b_r$, at a cost, $c_r<b_r$, whereas on poor patches the benefit is $b_p$ and the cost is $c_p<b_p$. Moreover, we assume $c_p>0>c_r$, i.e., cooperators exploit rich patches and provide benefits to their interaction partners including a personal gain of $-c_r$. This exploitation results in the degradation of rich patches. Conversely, on poor patches, cooperation bears a net cost but still comes at the expense of the environment in that the patch is unable to regenerate. The negative feedback between cooperators and rich patches results in interesting dynamics; see \fig{lambdarhocdeg}c. For fast ecological changes, $\lambda>\lambda_c=-\sigma c_r/c_p$, ${\bf P_0}$ is the only stable fixed point and defectors dominate with the exception of a small region of bi-stability between ${\bf P_0}$ and a stable mixture. Conversely, cooperation at ${\bf P_1}$ is never stable as long as $c_p>0$. Therefore, cooperators and defectors coexist when $\lambda<\lambda_c$. For faster regeneration rates (larger $\rho$), the interior fixed point $\bf Q$ marks stable coexistence, but for smaller $\rho$, $\bf Q$ also becomes unstable and gives rise to persistent oscillations resulting from stable limit cycles.

The limit $b_p,c_p\to0$ models opportunistic cooperators that follow utilitarian principles and cooperate only if occupying a rich patch, where their exploitation of the environment even provides personal benefits. For positive background fitness, $\sigma>0$, defection (on rich patches), ${\bf P_0}$, is never stable, while cooperation (on poor patches), ${\bf P_1}$, is always neutrally stable (largest eigenvalue of zero). No interior fixed point exists and eventually all trajectories converge to ${\bf P_1}$ and social interactions cease. An interesting twist arises for $\sigma=0$: now ${\bf P_0}$ is stable for sufficiently fast ecological changes, $\lambda>-c_r>0$, and neutrally stable otherwise, while the stability of ${\bf P_1}$ remains neutral. Moreover, a line of fixed points, $L$, appears and connects ${\bf P_0}$ with ${\bf P_1}$: $x_r=y_p=0$ and $x_p=1-y_r$. For $\lambda>-c_r$ all fixed points along $L$ are stable but for $\lambda<-c_r$ those with $x_p<\left(1+c_r/b_r-\sqrt{\left(1-c_r/b_r\right)^{2}-4\lambda/b_r}\right)\!\!/2$, i.e. closer to $\mathbf{P_0}$, are unstable. All equilibria reflect perfect segregation with all cooperators on poor patches and all defectors on rich ones. In the end, no social interactions take place because opportunistic cooperators on poor patches are just as passive as defectors on rich ones. The dynamics in this scenario are illustrated in \fig{lambdarhocdeg}d for small $b_p$ to avoid the intricacies of the degenerate case. $L$ is eliminated and either ${\bf P_0}$ or ${\bf P_1}$ is stable, depending on the rate of ecological changes.

\section{Discussion}
In nature, asymmetric interactions are the norm rather than the exception, whereas for models in evolutionary game theory the opposite holds. Even individuals of the same species tend to differ in more ways than in just some strategic behaviour, and these differences inevitably result in asymmetries. For example, individuals may differ in body size, strength or agility, experience different developmental histories or availability and access to resources -- all of which are likely to have some impact on the costs and benefits experienced by each individual in interactions. In general, such differences can be attributed either to the genetic makeup of an individual or its environmental conditions. Interestingly, asymmetries based on genetic (heritable) features can be easily dealt with by formally making the feature part of the strategic type of an individual. More specifically, $d$ strategies and $n$ genetic traits translate into $d n$ strategic types, and their interaction is represented by a \emph{symmetric} $d n\times d n$ payoff matrix \citep{mcavoy:PLoSCB:2015}, which provides ample justification for the historical focus on symmetric interactions.

In contrast, here we emphasize heterogeneities in the environment of the population by considering patches of different qualities that are occupied by one individual. This setup (\emph{i}) generates persistent asymmetric interactions between individuals, (\emph{ii}) explicitly models effects of patch quality, and (\emph{iii}) introduces feedback mechanisms between an individual's strategy and the quality of its patch. As a result, we observe rich eco-evolutionary dynamics solely based on ecological parameters given by the rate of degradation and restoration of patch qualities as well as the overall rate of change as compared to evolutionary changes. Indeed, fast ecological changes support cooperation in the asymmetric donation game with environmental benefits provided that cooperators are sufficiently adept at restoring poor patches. In contrast, if cooperators exploit their environment in order to confer benefits to their co-players, fast ecological changes may eliminate cooperation even in a harmony game, where cooperation is, in principle, the dominant and mutually-preferred strategy.

The type of feedback between strategies and patch quality can even introduce different interpretations of the nature of the traits themselves. While a cooperator may appear to act altruistically in social interactions by conferring benefits to their co-players, this same cooperator may exploit the resources of its patch and act as a cheater from the perspective of the environment. The tension between the socially beneficial and the environmentally beneficial strategies renders this case particularly interesting and defies classifications in terms of cooperators and defectors in the traditional sense.

This phenomenon is similar to but more striking than asymmetries in the snowdrift game where environmental effects may alter the characteristics of the interaction such that some individuals are effectively facing a stricter prisoner's dilemma or a harmony game that eliminates the social dilemma. Again, the interpretation of a seemingly-cooperative trait needs to be put into perspective with its environment.

While the asymmetric donation game is a straightforward extension of the symmetric donation game, there is no canonical way to account for asymmetry in the snowdrift game. Here, we have considered two biologically-reasonable versions of an asymmetric snowdrift game: one with equal splits and one with fair splits of the work between two cooperators. For equal splits, each cooperator clears half of the snowdrift (one at higher costs than the other), whereas for fair splits each bears the same costs (one clearing more than half of the snowdrift). Equality and fairness coincide in the symmetric snowdrift game since the costs for both players are the same.

For symmetric interactions, environmental feedback has recently been proposed \citep{weitz:PNAS:2016} as a way to resolve the tragedy of the commons \citep{hardin:Science:1968}. Their setup differs in that all individuals experience the same environment and requires that the characteristics of interactions change based on the condition of the environment. By construction, this model ensures that defection is dominant in rich environments, while cooperation dominates in poor environments. Not surprisingly, then, this model supports co-existence of cooperators and defectors in an intermediate environment or through cyclic dynamics. The impact of eco-evolutionary feedback has also been reviewed in the context of toxic algal blooms \citep{driscoll:EL:2015}, which reflects ecosystem engineering on microbial scales with macroscopic consequences. At low densities of toxic algae, the toxin may serve as a private good to deter predators or kill prey and hence selection favours toxin production. Conversely, at high densities, the increased toxin concentration may turn the hunter into the hunted and even invert trophic interactions if predators cannot tolerate high toxicity and become prey themselves. This turns toxin production into a public good where all algae benefit from the absence of predators or the provision of food resources. However, now selection acts against the costly production of toxins, which can give rise to cyclic dynamics. Our model, specifically the asymmetric donation game with negative environmental feedback, captures these dynamics, where rich (resp. poor) patches indicate the presence (resp. absence) of predators, c.f. \fig{lambdarhocdeg}c. Cooperation represents the costly production of toxins, which translates into immediate benefits in rich environments but results in net costs in poor environments.

Feedback between an individual's strategy and its environment introduces potential for linkage between strategic types and patch qualities. In the absence of feedback mechanisms, cooperators are doomed in prisoner's dilemma interactions but sufficiently strong positive feedback between cooperators and rich patches enables cooperators to persist or even take over the population. The emerging linkage represents a form of assortment \citep{fletcher:PRSB:2009}, only here cooperators do not offset their losses against defectors by more frequently interacting with other cooperators but rather by more frequently enjoying the benefits provided by rich patches.

\subsection*{Acknowledgments}
Financial support is acknowledged from the Natural Sciences and Engineering Research Council of Canada (NSERC), grant RGPIN-2015-05795 (C. H.), and the Office of Naval Research, grant N00014-16-1-2914 (A. M.).

\setcounter{section}{0}
\renewcommand{\thesection}{Appendix~\Alph{section}}
\renewcommand{\thesubsection}{\Alph{section}.\arabic{subsection}}
\renewcommand{\theequation}{\Alph{section}\arabic{equation}}
\renewcommand{\thefigure}{\Alph{section}\arabic{figure}}

\setcounter{equation}{0}
\setcounter{figure}{0}
\section{\label{app:dynnoalpha}No environmental feedback}
In the absence of environmental feedback, $\lambda=0$, \eq{dynxan} represents the transformed dynamics of \eq{dynnoalpha}. Naturally, the fraction of rich patches, $\alpha$, remains constant, $\dot\alpha=0$. Moreover, the linkage between cooperators and rich patches declines over time because $\dot\nu\propto-\nu$ and eventually disappears, see \eq{dynxan}c with $\lambda=0$. Thus, the long-term dynamics of the population are governed by \eq{dynxan}a in the limit $\nu\to0$: 
\begin{align}
\dot x &= x(1-x)\left[\alpha(\pi_1-\pi_3)+(1-\alpha)(\pi_2-\pi_4)\right]
\end{align}
with
\begin{subequations}
\label{eq:f}
\begin{align}
\pi_1&=\left({\bf x\cdot A}\right)_1 = x_r a_{11}+x_p a_{12}+y_r a_{13}+y_p a_{14} ; \\ 
\pi_2&=\left({\bf x\cdot A}\right)_2 = x_r a_{21}+x_p a_{22}+y_r a_{23}+y_p a_{24} ; \\ 
\pi_3&=\left({\bf x\cdot A}\right)_3 = x_r a_{31}+x_p a_{32}+y_r a_{33}+y_p a_{34} ; \\ 
\pi_4&=\left({\bf x\cdot A}\right)_4 = x_r a_{41}+x_p a_{42}+y_r a_{43}+y_p a_{44} ,
\end{align}
\end{subequations}
where ${\bf x}=(x_r, x_p, y_r, y_p)$ denotes the state of the population and the matrix ${\bf A} = \left[a_{ij}\right]$ denotes the payoff of an individual of type $i$ against one of type $j$ with $i,j\in\{C_r, C_p, D_r, D_p\}$. Note that \eq{dynxan}a is independent of $\sigma$ and hence the population dynamics remain unaffected by the choice of background fitness. Rearranging terms results in
\begin{align}
\label{eq:symavg}
\dot x &= x\left(1-x\right)\left( \overline{S}-\overline{P}-x \left(\overline{R}-\overline{S}-\overline{T}+\overline{P}\right) \right),
\end{align}
where $\overline{R}$, $\overline{S}$, $\overline{T}$, and $\overline{P}$ denote the payoffs averaged across the different patch qualities, i.e.
\begin{subequations}
\begin{align}
\overline{R} &=\alpha^2 a_{11}+\alpha\left(1-\alpha\right)\left(a_{12}+a_{21}\right)+\left(1-\alpha\right)^2 a_{22} ; \\ 
\overline{S} &=\alpha^2 a_{13}+\alpha\left(1-\alpha\right)\left(a_{14}+a_{23}\right)+\left(1-\alpha\right)^2 a_{24} ; \\ 
\overline{T} &=\alpha^2 a_{31}+\alpha\left(1-\alpha\right)\left(a_{32}+a_{41}\right)+\left(1-\alpha\right)^2 a_{42} ; \\
\overline{P} &=\alpha^2 a_{33}+\alpha\left(1-\alpha\right)\left(a_{34}+a_{43}\right)+\left(1-\alpha\right)^2 a_{44} .
\end{align}
\end{subequations}
Incidentally, \eq{symavg} is identical to the replicator equation for two types with payoff matrix
\begin{align}
\label{eq:abcdbar}
\bordermatrix{
& C & D \cr
C & \overline{R} & \overline{S} \cr
D & \overline{T} & \overline{P} \cr}.
\end{align}
Thus, the equilibria and hence the long-term dynamics of the \emph{asymmetric} interactions with four types are identical to an averaged \emph{symmetric} game between cooperators and defectors.

For asymmetric interactions captured by payoff matrices~(\ref{eq:abcde})-(\ref{eq:abcderp}) this yields 
\begin{align}
\label{eq:A}
\hspace{-3ex}{\bf A} &= \bordermatrix{
& C_r & C_p & D_r & D_p \cr
C_r & R+e & R+e & S+e & S+e \cr
C_p & R & R & S & S\cr
D_r & T+e & T+e & P+e & P+e \cr
D_p & T & T & P & P}
=
\begin{pmatrix}
R & R & S & S \\
R & R & S & S \\
T & T & P & P \\
T & T & P & P
\end{pmatrix}
+
\begin{pmatrix}
e & e & e & e \\
0 & 0 & 0 & 0 \\
e & e & e & e \\
0 & 0 & 0 & 0 
\end{pmatrix} ,
\end{align}
and formally highlights that, in this scenario, environmental benefits of a rich patch amount to a simple additive term to the fitness of its occupant. With $\sigma\geqslant -\min\{R,S,T,P\}$ the average fitness of cooperators and defectors are
\begin{subequations}
\label{eq:fcfdabcde}
\begin{align}
f_C &=x\left(\sigma+S +x\left(R-S\right)\right)+x_r e ; \\
f_D &=\left(1-x\right)\left(\sigma+P+x\left(T-P\right)\right)+y_r e ,
\end{align}
\end{subequations}
respectively. The long-term dynamics are determined by the average payoffs $\bar R=R+\alpha e$, $\bar S=S+\alpha e$, $\bar T=T+\alpha e$, and $\bar P=P+\alpha e$. Hence, in the absence of feedback mechanisms, additive environmental benefits merely result in a uniform shift of payoffs without changing the equilibria.

\setcounter{equation}{0}
\setcounter{figure}{0}
\section{\label{app:transform}Transforming the dynamics}
The dynamics of the four types with positive feedback between cooperators and rich patches, \eq{dynalpha}, can be transformed using more intuitive and illustrative quantities given by the frequency of cooperators, $x=x_r+x_p$, the frequency of rich patches, $\alpha=x_r+y_r$ and the linkage between cooperators and rich patches, $\nu=x_r-x \alpha$:
\begin{subequations}
\label{eq:dynxan}
\begin{align}
\dot x &=\ \dot x_r+\dot x_p = (1-x)f_C-x f_D\notag\\
&=\ x(1-\!x)\left[\alpha(\pi_1\!-\!\pi_3)+(1\!-\!\alpha)(\pi_2\!-\!\pi_4)\right]\notag\\
&\qquad +\nu\left[(1\!-\!x)(\pi_1\!-\!\pi_2)+x(\pi_3\!-\!\pi_4)\right] ; \\
\dot\alpha &=\ \dot x_r+\dot y_r = \lambda (x_p-\rho y_r)\notag\\
&=\ \lambda(x-\alpha\rho+(\rho-1)(\nu+\alpha x)) ; \\
\dot\nu &=\ \dot x_r-\alpha\dot x-x\dot\alpha = \notag\\
&= f_C y_r\!-\!f_D x_r\!+\!\lambda x_p-\alpha\left((1-x)f_C\!-\!x f_D\right)-x\lambda\left(x_p\!-\!\rho y_r\right)\notag\\
&=\ -\nu\big[\alpha \pi_3+(1\!-\!\alpha) \pi_4-x(\pi_4\!-\!\pi_2)\notag\\
&\qquad+(\nu+\alpha x)(\pi_1\!-\!\pi_2\!-\!\pi_3\!+\!\pi_4)+\lambda+\sigma\big]\notag\\
&\qquad +\lambda x[(\rho-1)(\alpha(1-x)-\nu)+1-x] .
\end{align}
\end{subequations}

For negative feedback between cooperators and rich patches, \eq{dyncdeg}, the dynamics can be similarly transformed:
\begin{subequations}
\label{eq:dynxneg}
\begin{align}
\dot x &=\ x(1-\!x)\left[\alpha(\pi_1\!-\!\pi_3)+(1\!-\!\alpha)(\pi_2\!-\!\pi_4)\right]\notag\\
&\qquad+\nu\left[(1\!-\!x)(\pi_1\!-\!\pi_2)+x(\pi_3\!-\!\pi_4)\right] ; \\
\dot\alpha &=\ \lambda(\rho(1-x)-\alpha\rho+(\rho-1)(\nu+\alpha x)) ; \\
\dot\nu &=\ -\nu\big[\alpha \pi_3+(1\!-\!\alpha) \pi_4-x(\pi_4\!-\!\pi_2)\notag\\
&\qquad+(\nu+\alpha x)(\pi_1\!-\!\pi_2\!-\!\pi_3\!+\!\pi_4)+\lambda+\sigma\big]\notag\\
&\qquad+\lambda x[(\rho-1)(\alpha(1-x)-\nu)-\rho(1-x)] .
\end{align}
\end{subequations}
Note that the evolutionary dynamics, Eq.~(\ref{eq:dynxneg}a), remains unchanged (c.f Eq.~(\ref{eq:dynxan}a)) but naturally the ecological dynamics and the linkage changes.

\setcounter{equation}{0}
\setcounter{figure}{0}
\section{\label{app:conservation}Conservation game}
\subsection{Stability analysis}
For the stability analysis in the conservation game, the Jacobian matrix, $\bf J$, is
\begin{align}
{\bf J} &= 
\small
\begin{pmatrix}
\left(2x-1\right)c & 0 & e \\
\lambda\left(1+\alpha\left(\rho-1\right)\right) & \lambda\left(x\left(\rho-1\right)-\rho\right) & \lambda\left(\rho-1\right) \\
\lambda\left(\substack{1+\nu\left(\rho-1\right) \\ +\left(2x-1\right)\left(\alpha\left(\rho-1\right)+1\right)}\right)+c\nu & \lambda x\left(1-x\right)\left(\rho-1\right)-e\nu & x\left(c-\lambda\left(\rho-1\right)\right)-\sigma-e \alpha -\lambda
\end{pmatrix}.
\end{align}
At the defector equilibrium, ${\bf P_0}$, the Jacobian, $\bf J$, simplifies to
\begin{align}
{\bf J} &= \begin{pmatrix}
-c & 0 & e\\
\lambda & -\lambda \rho & \lambda\left(\rho-1\right)\\
\lambda & 0 & -\sigma -\lambda
\end{pmatrix} ,
\end{align}
with eigenvalues $\xi_0=-\lambda\rho$ and $\xi_\pm=-\frac{1}{2}\left(\lambda+\sigma+c\pm\sqrt{\left(\lambda+\sigma-c\right)^2+4e\lambda}\right)$. All three eigenvalues are always real and negative as long as $\left(e-c\right)\lambda<c\sigma$. Hence ${\bf P_0}$ is a stable node but turns into an unstable node for small costs, $c$, large environmental benefits, $e$, or sufficiently fast ecological dynamics, $\lambda>\lambda_c=c\sigma/\left(e-c\right)$.

Similarly, at the cooperator equilibrium, ${\bf P_1}$, the Jacobian, $\bf J$, simplifies to
\begin{align}
{\bf J} &= \begin{pmatrix}
c & 0 & e\\
\lambda\rho & -\lambda & \lambda\left(\rho-1\right)\\
-\lambda\rho & 0 & -e+c-\sigma -\lambda\rho
\end{pmatrix}
\end{align}
with eigenvalues $\xi_0=-\lambda$ and $\xi_{\pm}=-\frac{1}{2}\left(\lambda\rho+\sigma+e-2c\pm\sqrt{\left(\lambda\rho+\sigma+e\right)^2-4e\lambda\rho}\right)$. Hence, if the inequality $\left(e+\sigma+\lambda\rho\right)^2>4\lambda\rho e$ holds, then all eigenvalues are real and ${\bf P_1}$ is a stable node provided $\lambda\rho\left(e-c\right)>c\left(e-c+\sigma\right)$. Conversely, if $\left(e+\sigma+\lambda\rho\right)^2<4\lambda\rho e$, then two eigenvalues are complex conjugates and ${\bf P_1}$ is a stable focus whenever $e+\sigma-2c+\lambda\rho<0$, which requires $e<c$ because $\sigma>c$.

Unfortunately, at ${\bf Q}$ the Jacobian has an irreducible cubic characteristic polynomial, which essentially renders any further analytical investigations unfeasible, except in special cases based on further simplifying assumptions.

\subsection{\label{app:hopf}Degradation equals restoration, $\rho=1$}
For equal degradation and restoration rates, $\rho=1$, the interior fixed point is $\textbf{Q}=\left(\hat x, \hat x, \left(1-\hat x\right)\hat x c/e\right)$ with $\hat x=\lambda/c-\sigma/\left(e-c\right)$ and the Jacobian simplifies to
\begin{align}
{\bf J}\! &=\! \small\begin{pmatrix}
\left(2\hat x-1\right)c & 0 & e\\
\lambda & -\lambda & 0\\
\lambda\left(1\!-2\hat x\right)+\frac{c^2}{e}(1\!-\hat x)\hat x\!\!\! & -(1-\hat x)\hat x c\!\!\! & \hat x(c-e)-\sigma\!-\!\lambda
\end{pmatrix}.
\end{align}
In this case, it is possible to determine whether a Hopf bifurcation may occur, and hence check the potential for stable limit cycles. One can do so by considering the characteristic polynomial for $\bf J$ in the form $\xi^3+p\xi^2+q\xi+r=0$, where the roots $\xi$ represent the eigenvalues of $\bf J$. If a Hopf bifurcation occurs, then a parameter combination must exist such that one eigenvalue is real, $\xi_0$, and the other two are a purely imaginary complex conjugate pair, $\pm i\xi_1$, which results in the factorization $\left(\xi-\xi_0\right)\left(\xi^2+\xi_1^2\right)=\xi^3-\xi_0\xi^2+\xi_1^2\xi-\xi_0\xi_1^2=0$. Hence, at this point, the polynomial coefficients need to satisfy $p=-\xi_0, q=\xi_1^2$ and $r=-\xi_0\xi_1^2$ or $r=pq$, thus
\begin{align}
pq-r = e^{3}\left(e-c\right)^{3}c \Big[ &e\left(e-c\right)^{4}\lambda^{3}+\left(e-c\right)^{2}c^{3}\left(c^{2}+2\lambda\left(e-c\right)\right)\sigma \nonumber \\
&\quad +\left(e-c\right)c^{3}\left(3c^{2}+2\lambda\left(e-c\right)\right)\sigma^{2}+2c^{5}\sigma^{3} \Big].
\end{align}
For $e>c$, all terms on the right-hand-side are positive and hence the conditions for a Hopf bifurcation are never met.

\bibliographystyle{unsrtnat}

\end{document}